# Faster Approximation Schemes for Fractional Multicommodity Flow Problems via Dynamic Graph Algorithms

Aleksander Mądry[*]

Massachusetts Institute of Technology

madry@mit.edu

## Abstract

We combine the work of Garg and Könemann, and Fleischer with ideas from dynamic graph algorithms to obtain faster $(1 - \varepsilon)$-approximation schemes for various versions of the multicommodity flow problem. In particular, if $\varepsilon$ is moderately small and the size of every number used in the input instance is polynomially bounded, the running times of our algorithms match – up to poly-logarithmic factors and some provably optimal terms – the $\Omega(mn)$ *flow-decomposition barrier* for *single-commodity* flow.

## 1 Introduction

Flow problems are one of the most fundamental problems in optimization; they have a long list of scientific and engineering applications — see e.g. Ahuja *et al.* [1]. Usually, an instance of such problem consists of a directed graph $G = (V, E, u)$ with capacities $u : E \to \mathbb{R}^+$ on the arcs, and a source-sink pair $(s_1, t_1)$ (in case of single-commodity flow); or a set of $k$ source-sink pairs $\{(s_i, t_i)\}_{1 \le i \le k}$ (in case of multicommodity flow). The task is to find flows $(f_1, \ldots, f_k)$, where $f_i$ is a flow of commodity $i$ from $s_i$ to $t_i$, that optimize some objective function while each flow satisfies node conservation constraints and the total flow $\sum_i f_i(e)$ of the commodities through any arc $e$ is not exceeding its capacity $u(e)$. The simplest multicommodity flow problem is the *maximum multicommodity flow* problem – in this case the objective is to maximize the sum of the flows $\sum_i |f_i|$, where $|f_i|$ is the amount of commodity $i$ routed from $s_i$ to $t_i$. A generalization of this problem is the *maximum weighted multicommodity flow* problem in which additionally we are given weights $w_1, \ldots, w_k$ and we want to maximize the weighted sum of the flows $\sum_i w_i |f_i|$. Another popular variation of the multicommodity flow problem is the *maximum concurrent flow* problem. In this problem, we are given a set of $k$ positive demands $d_1, \ldots, d_k$ and are asked to find a multicommodity flow that is feasible (i.e. obeys arc capacities) and routes $\lambda d_i$ units of commodity $i$ between each source-sink pair $(s_i, t_i)$ – the goal is to maximize the value of $\lambda$. If there is a cost function $c : E \to \mathbb{R}^+$ associated with arcs, where $c(e)$ is the price of routing one unit of flow through arc $e$, the *minimum cost concurrent flow* problem is to find a maximum concurrent flow whose total cost i.e. the sum of the costs incurred by the flow on each arc, is within some target budget $B$.

---

[*]Supported by a Fulbright Science and Technology Award, by NSF contract CCF-0829878, and by ONR grant N00014-05-1-0148

Although all the problems defined above can be solved optimally in polynomial time by formulating them as linear programs, in many applications it is more important to compute an approximate solution fast than to compute an optimal one. Therefore, much effort was put into obtaining efficient *fully polynomial-time approximation schemes* (FPTAS) for multicommodity flow problems. A fully polynomial-time approximation scheme for a maximization problem is an algorithm that, given an instance of the problem and an accuracy parameter $\varepsilon > 0$, computes, in time polynomial in the size of the input and $1/\varepsilon$, a solution that has objective value within a factor of $(1 - \varepsilon)$ of the optimal one.

## 1.1 Previous work

Over the past two decades there has been a rich history of results providing FPTASes for multicommodity flow problems. Shahrokhi and Matula [30] presented the first combinatorial fully polynomial-time approximation scheme for the maximum concurrent flow problem with uniform arc capacities, and introduced the idea of using an exponential length function to control arc congestion. Subsequently, a series of results [19, 21, 13, 15, 24, 31, 25, 18, 16] based on Langrangian relaxation and linear programming decomposition yielded algorithms that had significantly improved running time and could be applied to various versions of the multicommodity flow problem with arbitrary arc capacities. All the above algorithms compute an initial (infeasible) flow and then redistribute it from more congested paths to less congested ones by repeatedly solving an oracle subproblem of either minimum cost single-commodity flow [21, 13, 15, 25, 18, 16], or shortest path [30, 19, 24, 31].

In [32], Young deviated from this theme by presenting an *oblivious rounding* algorithm that avoids rerouting of the flow. Instead, it builds the solution from scratch. At each step it employs shortest path computations (with respect to exponential length function that models the congestion of the arcs) to augment the flow along suitable (i.e. relatively uncongested) paths. At the end, it obtains the final feasible solution by scaling down the flow by the maximum congestion it incurred on arcs. A similar approach was taken by Garg and Könemann [12]; however they managed to provide an elegant framework for solving multicommodity flow problems that yields a simple analysis of the correctness of the obtained algorithms. This allowed them to match and, in some cases, improve over the running time of the algorithms obtained via the redistribution methodology. Subsequently, Fleischer [11] used this framework to develop significantly faster algorithms for various multicommodity flow problems. In particular, for the maximum multicommodity flow problem she managed to obtain a running time of $\widetilde{O}(m^2 \varepsilon^{-2})$ (where $\widetilde{O}(\cdot)$ notation hides poly-logarithmic factors) that is independent of the number of commodities. For the weighted version, she proposed an algorithm running in $\widetilde{O}(m^2 \varepsilon^{-2} \min\{\log M, k\})$ time, where $\log M$ is the upper bound on the size of binary representation of every number used in the input instance. For the maximum (resp. minimum cost) concurrent flow problem her algorithm has a running time of $\widetilde{O}((m+k)m\varepsilon^{-2})$ (resp. $\widetilde{O}((m + k)m\varepsilon^{-2} \log M)$). Her results for both versions of the concurrent flow problem were later improved by Karakostas [17], who was able to reduce the term in the running time that depends on $k$ from $\widetilde{O}(km\varepsilon^{-2})$ (resp. $\widetilde{O}(km\varepsilon^{-2} \log M)$) to $\widetilde{O}(kn\varepsilon^{-2})$. Interestingly, he also showed that if we want to obtain the $(1 - \varepsilon)$-approximation of only the *value* of the maximum (resp. minimum cost) concurrent flow rate (without obtaining the actual flow) then this can be done in $\widetilde{O}(m^2 \varepsilon^{-2})$ (resp. $\widetilde{O}(m^2 \varepsilon^{-2} \log M)$) time.

All the above algorithms have quadratic dependence of their running times on $1/\varepsilon$. Klein and Young [20] give evidence that this quadratic dependence is inherent for *Dantzig-Wolfe-type* algorithms[1] — all the FPTASes mentioned so far are of this type. As it turns out, better dependence on $1/\varepsilon$ can be obtained. Bienstock and Iyengar [6] adapted the technique of Nesterov [22] to give an FPTAS that has $O(\frac{1}{\varepsilon \log 1/\varepsilon})$ dependence on $1/\varepsilon$. Very recently, Nesterov [23] obtained an approximation scheme where this dependence is just linear. However, the running times of both these algorithms have worse dependence on parameters other than $1/\varepsilon$ compared to the algorithms described above – for example, the approximation scheme due to Nesterov has the running time of $\widetilde{O}(k^2 m^2 \varepsilon^{-1})$.

## 1.2 Our contribution

We build on the work of Garg and Könemann [12] and Fleischer [11] to obtain faster approximation schemes for various versions of multicommodity flow problems. At a high level — when the size of every number used in the input instance (e.g. capacities, weights, and costs on arcs) is polynomially bounded — our improvements break the bottlenecking term of $\Omega(m^2 \varepsilon^{-2})$ that all the previous Dantzig-Wolfe-type algorithms suffer from, by substituting it with $\widetilde{O}(mn\varepsilon^{-2})$ term. Our result is based on two main ideas.

The first one stems from an observation that the shortest-path subproblems that algorithms following the Garg-Könemann framework solve repeatably are closely related. Namely, each successive subproblem corresponds to the same underlying graph – only the lengths of some of the arcs are increased. This suggests that treating each of these subproblems as an independent task – as it is the case in all the previous algorithms – is suboptimal. One might wonder, for example, whether it is possible to maintain a data structure that allows answering such a sequence of shortest-path queries more efficiently than just by computing everything from scratch in each iteration. Indeed, it turns out that this kind of questions were already studied extensively in the area of *dynamic graph algorithms* (see e.g. [10, 3, 7, 9, 26, 4, 8, 27, 29, 28, 5]). In particular, the data structure that we would like to maintain corresponds to the *decremental dynamic all-pair shortest path* problem. Unfortunately, if we are interested in solutions whose overall running time is within our intended bounds, then it seems there is no suitable existing result that can be used (see section 3 for details).

This lack of existing solution fitting our needs brings us to the second idea of the paper. We note that when we employ the Garg-Könemann framework to solve multicommodity flow problems, it is not necessary to compute the (approximately) shortest path for each shortest-path subproblem. All we really need is that the set of the suitable paths over which we are optimizing the length comes from a set that contains *all* the flowpaths of some *fixed* optimal solution to the multicommodity flow instance that we are solving. To exploit this fact, we introduce a random set of paths $\widehat{\mathcal{P}}$ (see Definition 5) – that can be seen as a sparsification of the set of all paths in $G$, and that with high probability has the above-mentioned property. Next, we combine the ideas from dynamic graph algorithms to design an efficient data structure that maintains all-pair shortest path distances with respect to the set $\widehat{\mathcal{P}}$. This data structure allows us to modify (in an almost generic manner) the existing algorithms for various multicommodity flow problems that are based on the Garg-Könemann framework and transform them into Monte-Carlo algorithms with improved running times.

The summary of our results and their comparison with the previous ones can be found in Figure 1. Note that there exist instances of concurrent flow problems (see Figure 2) for which any (approximately) optimal solution has representation size of $\Omega(kn)$. Therefore, in the setting where $\varepsilon$ is fixed or moderately small, say $1/\log^{O(1)} n$, and the size of every number used in the input instance is polynomially bounded, the corresponding $\widetilde{O}(kn)$ term in the running time of our algorithms for concurrent flow problems is optimal (up to poly-logarithmic factors). Interestingly, in this setting, the running time of all our algorithms matches – up to poly-logarithmic factor and modulo the nearly-optimal $\widetilde{O}(kn)$ term in case of concurrent flow problems – the $\Omega(mn)$ *flow-decomposition barrier* for *single-commodity* flows. Recall that in the case of the maximum single-commodity flow problem Goldberg and Rao [14] presented an algorithm that improves upon this barrier. Thus we find the question whether one can also achieve a similar improvement for the maximum multicommodity flow problem very intriguing.[2]

---

[1]A *Dantzig-Wolfe-type* algorithm for a fractional packing problem – in which the goal is to find $x$ in some polytope $P$ that satisfies the set of packing inequalities $Ax \leq b$ – is an algorithm that accesses $P$ only by queries of the form: "given a vector $c$, what is the $x \in P$ minimizing $c \cdot x$?".

| Problem | Previous best | This paper |
|---|---|---|
| maximum multicommodity flow | $\widetilde{O}(m^2\varepsilon^{-2})$ [11] | $\widetilde{O}(mn\varepsilon^{-2})$ |
| maximum weighted multicommodity flow | $\widetilde{O}(m^2\varepsilon^{-2}\min\{\log M,k\})$ [11] | $\widetilde{O}(mn\varepsilon^{-2}\log^2 M)$ |
| maximum concurrent flow | $\widetilde{O}((m^2+kn)\varepsilon^{-2})$ [17] | $\widetilde{O}((m+k)n\varepsilon^{-2}\log M)$ |
| | $\widetilde{O}(k^2m^2\varepsilon^{-1})$ [23] | |
| minimum cost concurrent flow | $\widetilde{O}((m^2\log M+kn)\varepsilon^{-2})$ [17] | $\widetilde{O}((m+k)n\varepsilon^{-2}\log M)$ |

Figure 1: Comparison of $(1-\varepsilon)$-approximation schemes for multicommodity flow problems. $\widetilde{O}(\cdot)$ notation hides poly-logarithmic factors, $m$ is the number of arcs, $n$ is the number of vertices, $k$ is the number of commodities, and $\log M$ is the upper bound on the size of binary representation of any number used in the input instance.

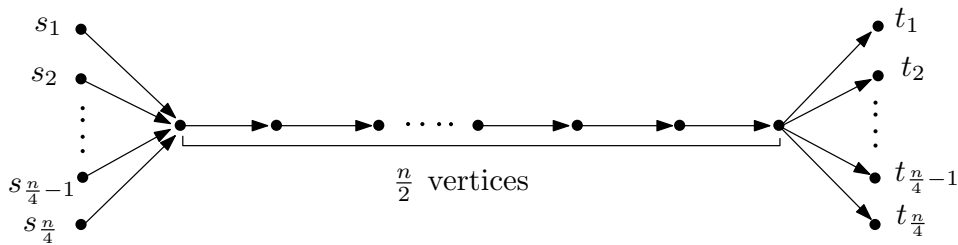

Figure 2: By choosing any set of $k \le n/16$ distinct $(s_i, t_j)$ pairs as source-sink pairs for $k$ commodities and setting all demands to 1, we obtain an instance of the maximum concurrent flow problem for which any $(1-\varepsilon)$-approximate solution has representation size $\Omega(kn)$ – each of $n/2-1$ arcs on the middle path has to have non-zero flow of all $k$ commodities flowing through it.

---

[2]Note that the only obvious lower bound for the running time of an algorithm solving the maximum multicommodity problem is $\Omega(m+\min\{k,n\}n)$. This bound can be established by constructing instances as in Figure 2, where each of $k \le n$ commodities corresponds to the source-sink pair $(s_i, t_i)$ for $1 \le i \le k$, all the arcs outgoing from $s_i$s have capacity of 1, and the rest of the arcs have capacity of $k$. Clearly, any (approximately optimal) solution to such instances has representation size of $\Omega(\min\{k,n\}n)$.

### 1.3 Notations and Definitions

Let $G = (V, E, u)$ be a *directed* graph with capacities $u : E \to \mathbb{R}^+$. In addition to capacities, we will often equip arcs of $G$ with lengths given by some *length function* $l : E \to \mathbb{R}^+$. For any directed path $p$ in $G$, by the *length of $p$ with respect to $l$* we mean a quantity $l(p) := \sum_{e \in p} l(e)$. For any two vertices $u$ and $v$ of $G$, a $u$-$v$ path is a directed path in $G$ that starts at $u$ and ends at $v$. We define *distance from $u$ to $v$ (with respect to $l$)* for $u, v \in V$ to be the length (with respect to $l$) of the shortest $u$-$v$ path. We will omit the reference to the length function $l$ whenever it is clear from the context which length function we are using.

For $1 \le i \le k$, we denote by $\mathcal{P}_i$ the set of all $s_i$-$t_i$ paths in $G$, where $\{(s_i, t_i)\}_i$ is the set of $k$ source-sink pairs of the instance of the multicommodity flow problem we are considering. Let $\mathcal{P} = \bigcup_{i=1}^{k} \mathcal{P}_i$. For a given subset $U \subseteq V$ of vertices, let $\mathcal{P}(U)$ be the set of all paths in $\mathcal{P}$ that pass through at least one vertex from $U$. Finally, for a given $j > 0$, let $\mathcal{P}(U, j)$ be the set of all the paths from $\mathcal{P}(U)$ that consist of at most $j$ arcs.

### 1.4 Outline of the paper

We start with section 2 where we illustrate the Garg-Könemann framework [12] for the maximum multicommodity flow problem – we also outline the current best algorithm for this problem due to Fleischer [11]. Next, in section 3, we introduce the main ideas and tools behind our results – the connection between fast approximation schemes for multicommodity flow problems and dynamic graph algorithm for maintaining (approximately) shortest paths with respect to the sparsified set of paths. In particular, we formally define the set $\widehat{\mathcal{P}}$, and the $(\delta, M_{\max}, M_{\min}, \widehat{\mathcal{P}})$-ADSP data structure that we will be using for maintenance of the (approximately) shortest paths that we are interested in. Subsequently, in section 4, we show how these concepts lead to a more efficient algorithm for the maximum multicommodity flow problem, and then extend it to handle the weighted version of the problem. In section 5, we obtain an improved algorithm for the maximum concurrent flow problem and outline its generalization to the version with costs. We conclude in section 6 with an efficient implementation of the $(\delta, M_{\max}, M_{\min}, \widehat{\mathcal{P}})$-ADSP data structure.

## 2 Garg-Könemann framework for solving multicommodity flow problems

Our algorithms will follow the framework for solving multicommodity flow problems that was developed by Garg and Könemann [12] (see also [2] for a presentation of this framework from a slightly different perspective). For illustrative purposes, we now focus only on the variation of the framework for the *maximum multicommodity flow* – later, we will describe the variations corresponding to other types of multicommodity flow problems.

The starting point of this framework is the following path-based linear programming formulation of the maximum multicommodity problem:

$$\max \sum_{p \in \mathcal{P}} f_p$$

$$\text{s.t.} \sum_{p \ni e} f_p \leq u(e) \qquad\qquad \forall e \in E, \tag{1}$$

$$f_p \geq 0 \qquad\qquad \forall p \in \mathcal{P}.$$

Here $f_p$ represents the flow on path $p \in \mathcal{P}$, and we recall that $\mathcal{P}$ is the set of all $s_i$-$t_i$ paths in $G$, where $\{(s_i, t_i)\}_i$ is the set of $k$ source-sink pairs of the instance of the maximum multicommodity flow problem we are considering. The dual of this linear program corresponds to assigning lengths $l(\cdot)$ to the arcs in such a way that length of every path in $\mathcal{P}$ is at least one and the total volume of the network i.e. $\sum_{e \in E} l(e)u(e)$ is minimized.

$$\min \sum_{e \in E} l(e)u(e)$$

$$\text{s.t.} \sum_{e \in p} l(e) \geq 1 \qquad\qquad \forall p \in \mathcal{P}, \tag{2}$$

$$l(e) \geq 0 \qquad\qquad \forall e \in E.$$

Intuitively, the total volume is an upper bound on the value of the maximum multicommodity flow, since when lengths $l(\cdot)$ constitute a feasible dual solution, routing one unit of flow between any source-sink pair uses up a volume of at least one.

---

**Input**   : Graph $G = (V, E)$, capacities $u(e)$, commodity pairs $\{(s_i, t_i)\}_i$, accuracy parameter $\varepsilon > 0$
**Output**: (Infeasible) flow $f$

Initialize $f \leftarrow \emptyset$, and $l(e) \leftarrow \gamma$ for all arcs $e \in E$, where $\gamma = (1+\varepsilon)/((1+\varepsilon)n)^{1/\varepsilon}$
**while** there is a path $p \in \mathcal{P}$ with $l(p) < 1$ **do**

    Select path $p$ that is the shortest one (with respect to $l$) in $\mathcal{P}$
    Find the bottleneck capacity $u$ of $p$             $(*\ u \leftarrow \min_{e \in p} u(e)\ *)$
    Augment the flow $f$ by routing $u$ units of flow along the path $p$
    **foreach** arc $e$ in $p$ **do** $l(e) \leftarrow l(e)(1 + \frac{\varepsilon u}{u(e)})$

**end**
**return** $f$

Figure 3: Garg-Könemann algorithm for finding maximum multicommodity flow

---

To find a $(1 - \varepsilon)$-approximate solution to the above linear program (1) and thus to obtain corresponding multicommodity flow, Garg and Könemann employ an algorithm presented in Figure 3. The algorithm maintains flow $f$, and a length function $l$. Initially, $f = 0$, and $l = \gamma$ i.e. there is no flow routed, and the length of each arc is $\gamma$, where $\gamma = (1+\varepsilon)/((1+\varepsilon)n)^{1/\varepsilon}$ is very small. Now, as long as there are paths in $\mathcal{P}$ having length smaller than one, the algorithm chooses the shortest path $p$ among these paths and augments the flow $f$ along it. The amount of flow routed over $p$ is equal to the *bottleneck capacity* $u$ of $p$ i.e. $u$ is the minimal capacity among all the capacities of the arcs of $p$. After augmenting the flow, we update the length function by multiplying the length $l(e)$ of each arc $e$ of $p$ by a factor of $(1 + \frac{\varepsilon u}{u(e)})$.

It is not hard to see that the final flow $f$ produced by the above procedure may violate some of the arc capacities. Therefore, to obtain a feasible solution we need to scale the final flow $f$ down by the maximum congestion $f$ incurred on arcs. Since we only augment the flow along paths with length smaller than one and our length update rule ensures that the length of arcs is exponential in their congestion, we can conclude that this maximum congestion is not very large.

**Lemma 1** (see [12]). *The flow obtained by scaling the final flow $f$ down by $\log_{1+\varepsilon} \frac{1+\varepsilon}{\gamma}$ is feasible.*

Moreover, the fact that we always augment the flow along the shortest path in $\mathcal{P}$ – which intuitively means that we aim at routing the flow through arcs that are relatively uncongested – allowed Garg and Könemann to bound the quality of the obtained solution.

**Lemma 2** (see [12]). *If $f$ is the final flow computed then $\frac{|f|}{\log_{1+\varepsilon} \frac{1+\varepsilon}{\gamma}} \geq (1-2\varepsilon)OPT$, where $OPT$ is the value of the optimal flow.*

The key ingredient in the proof of the above lemma is an observation – a simple consequence of weak duality between linear programs (1) and (2) – that for *any* length function $l$ there is always a path $p$ in $\mathcal{P}$ whose length with respect to $l$ is at most $1/OPT$ fraction of the total volume of the graph $G$ i.e. $l(p) \leq \frac{\sum_e l(e)u(e)}{OPT}$.

Now, to analyze the running time of the algorithm, we note that in each augmentation the length of the arc with bottleneck capacity increases by a factor of $(1+\varepsilon)$. Since no arc achieves a length bigger than $(1+\varepsilon)$, there can be at most $m\lfloor \log_{1+\varepsilon} \frac{1+\varepsilon}{\gamma} \rfloor = \widetilde{O}(m\varepsilon^{-2})$ augmentations of the flow. To perform each such augmentation we have to find the shortest path in $\mathcal{P}$, this can be done in $\widetilde{O}(km)$ time by running Dijkstra's algorithm from each possible source. This establishes the following theorem.

**Theorem 3** (see [12]). *For any $\varepsilon > 0$ one can compute $(1-2\varepsilon)$-approximation to maximum multicommodity flow problem in time $\widetilde{O}(km^2\varepsilon^{-2})$.*

Subsequently, Fleischer [11] presented a more efficient version of the above algorithm. Her improvement is based on the realization that whenever we need the shortest path in the Garg-Könemann algorithm, it is sufficient to compute an $(1+\varepsilon)$-*approximately* shortest path. This allows for modification of the algorithm of Garg and Könemann to avoid solving the shortest-path problem for all the source-sink pairs each time the flow is augmented. Instead, we cycle through all commodities, keeping augmenting the flow along the shortest path corresponding to given commodity as long as the length of this path is at most $(1+\varepsilon)\widehat{\alpha}$ – where $\widehat{\alpha}$ is a lower bound estimate of the current length of the shortest path in $\mathcal{P}$ maintained by the algorithm – and moving on once it does not. To start, we set $\widehat{\alpha} = \gamma$ and we do not increase its value as long as we manage to augment the flow along some path in $\mathcal{P}$ of small enough length (i.e. at most $(1+\varepsilon)\widehat{\alpha}$). Once we are unable to find such a path i.e. our cycling through commodities made a full cycle, we set $\widehat{\alpha} \leftarrow (1+\varepsilon)\widehat{\alpha}$, and start cycling again unless $\widehat{\alpha}$ is already bigger than one – in which case we scale the obtained flow down to make it feasible and the algorithm terminates. The algorithm is presented in Figure 4. An important implementation detail is that when cycling through commodities, we group together source-sink pairs that share the same source. This allows us to take advantage of the fact that one execution of Dijkstra's algorithm computes simultaneously the shortest paths for all these pairs. To see why the above modifications reduce the number of shortest path computations needed, notice that each execution of Dijkstra's algorithm either results in flow augmentation, or causes us to move on in our cycling to the next group of commodities that was not yet examined. Note that our way of updating the value of $\widehat{\alpha}$ ensures that there is at most $\lfloor \log_{(1+\varepsilon)} \frac{1}{\gamma} \rfloor = O(\log m\varepsilon^{-2})$ full cycles through commodities. Also, the number of flow augmentation is still at most $m\lfloor \log_{1+\varepsilon} \frac{1+\varepsilon}{\gamma} \rfloor = \widetilde{O}(m\varepsilon^{-2})$. Therefore, we have at most $\widetilde{O}((m + \min\{k, n\})\varepsilon^{-2})$ executions of Dijkstra's algorithm – instead of $\widetilde{O}(km\varepsilon^{-2})$ executions in the original algorithm due to Garg and Könemann. Fleischer proves the following theorem.

**Theorem 4** (see [11]). *For any $0.15 > \varepsilon > 0$ the algorithm in Figure 4 runs in time $\widetilde{O}(m^2\varepsilon^{-2})$, and returns a flow that has value at least $(1 - 4\varepsilon)OPT$.*

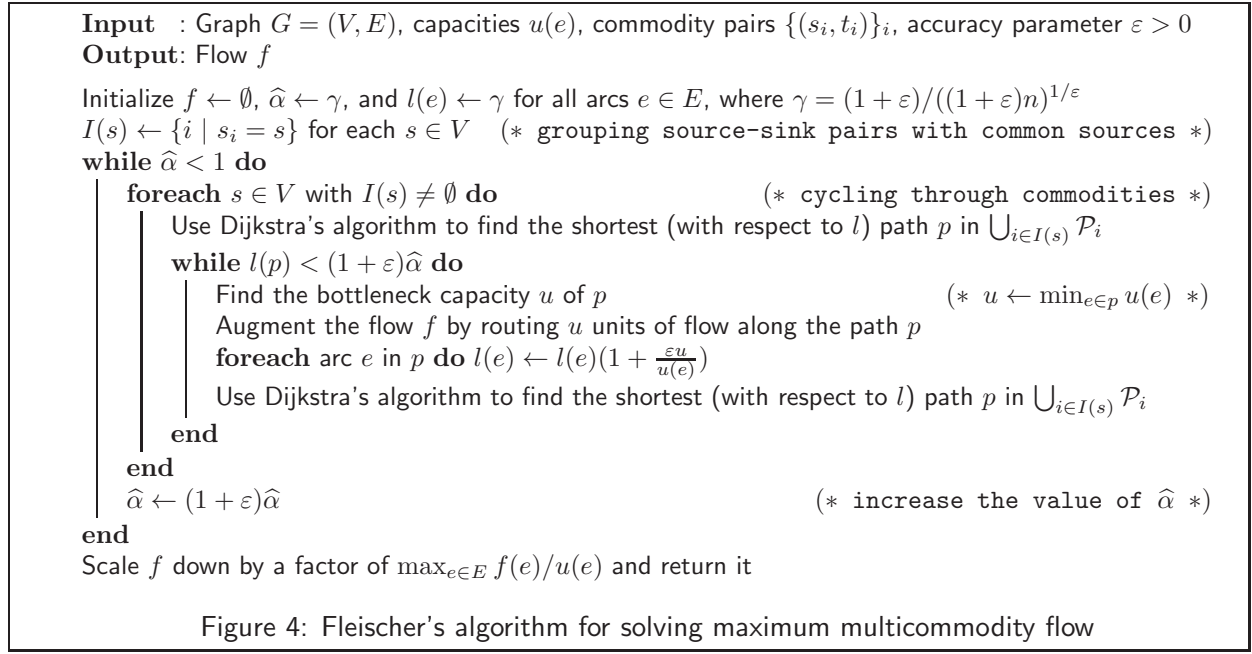

Figure 4: Fleischer's algorithm for solving maximum multicommodity flow

# 3  Solving multicommodity flow problems and dynamic graph algorithms

As described in the previous section, the approximation scheme for the maximum multicommodity flow problem due to Garg and Könemann [12] finds a near-optimal solution by repeatedly solving the oracle subproblem of computing the shortest path in graph $G$ with respect to some evolving length function $l$. (Later we will see that the same general approach is also used for the other variants of multicommodity flow problems.) As a consequence, the running time of the resulting approximation schemes is dominated by the time needed to solve these subproblems using Dijkstra's algorithm. By careful choice of the subproblems as well as better utilization of the computed answers, the number of these (single-source) shortest path computations was reduced considerably in subsequent improvements due to Fleischer [11] and Karakostas [17]. However, the reduced numbers are still $\Omega(m\varepsilon^{-2})$ which leads to a time complexity of $\Omega(m^2\varepsilon^{-2})$ for the corresponding algorithms.

The main observation behind the construction of our improved approximation schemes is that treating each of the oracle subproblems as an independent task (and thus using Dijkstra's algorithm each time), as the above-mentioned results do, is suboptimal. After all, each successive subproblem corresponds to the same underlying graph – only with lengths of some of the arcs increased. Therefore, one might hope to construct a data structure that solves such a sequence of subproblems more efficiently than by computing each time everything from scratch. More precisely, one might wonder whether there is an efficient data structure that maintains a directed graph $G$ with lengths on arcs and supports operations of: increasing a length of some arc; answering shortest-path distance query; and returning shortest vertex-to-vertex path.

It turns out that the problem of designing such a data structure is already known in the literature as the *decremental dynamic all-pairs shortest path* problem and extensive work have been done on this and related problems (see e.g. [10, 3, 7, 9, 26, 4, 8, 27, 29, 28, 5]). However, if we are interested in solutions whose overall running time is within our intended bounds, the result that is closest to what we need is the one by Roditty and Zwick [27]. They show if $G$ were *undirected* and had positive, *integer* lengths on edges, with maximum length being $b$, a $(1 + \delta)$-approximate solution to the decremental dynamic all-pair shortest path problem can be implemented with total maintenance time of $\widetilde{O}(\frac{mnb}{\delta})$, $O(1)$ time needed to answer any vertex-to-vertex shortest-path distance query, and returning shortest vertex-to-vertex path in $O(n)$ time.[3] Unfortunately, in our applications the graph $G$ is not only directed, but also the lengths of the arcs (when we scale them to make them integral) can be of order of $b = \Omega(n^{1/\varepsilon})$ with $\varepsilon < 0.15$. Therefore, the resulting running time would be prohibitive. Further, the construction of Roditty and Zwick assumes that the sequence of operations to be handled is oblivious to the behavior of the data structure (e.g. to its randomized choices). This feature is a shortcoming from our point of view since in our setting the way the length function of the graph evolves depends directly on which shortest paths the data structure chose to output previously.

To circumvent this lack of suitable existing solutions, we realize that for our purposes it is sufficient to solve a simpler task than the decremental dynamic all-pairs shortest path problem in its full generality (i.e. in the directed setting, and allowing large lengths on arcs and adversarial requests). Namely, when we are using the Garg-Könemann framework to solve multicommodity flow problems, it is not necessary to compute for each subproblem the (approximately) shortest path among all the suitable paths in the set $\mathcal{P}$. As we will prove later, to establish satisfactory bounds on the quality of the final flow, it suffices that whenever we solve some shortest-path subproblem, the set of suitable paths over which we are optimizing the length comes from a set that contains *all* the flowpaths of some fixed optimal solution to the instance of the multicommodity flow problem that we are solving.

With this goal in mind, we define the following random subset $\widehat{\mathcal{P}}$ of paths in $\mathcal{P}$.[4] One may view $\widehat{\mathcal{P}}$ as a sparsification of the set $\mathcal{P}$.

**Definition 5.** *For $j = 1, \dots, \lceil \log n \rceil$, let $S_j$ be a random set obtained by sampling each vertex of $V$ with probability $p_j = \min\{\frac{10 \ln n}{2^j}, 1\}$. Define $\widehat{\mathcal{P}} := \bigcup_{j=1}^{\lceil \log n \rceil} \mathcal{P}(S_j, 2^j)$, where $\mathcal{P}(U, j')$ for a given $U \subseteq V$, and $j' > 0$ is the set of all paths in $\mathcal{P}$ that consist of at most $j'$ arcs and pass through at least one vertex of $U$.*

---

[3] Note that there can be as many as $mb$ edge length increases in a sequence, thus this solution is faster than naïve solution that just computes all-pairs shortest-path after each edge-length increase, or one that just answers each of the queries using Dijkstra's algorithm.

[4] It is worth noting that a very similar set was used in [27] albeit with a slightly different motivation.

It turns out that this sparsification $\widehat{\mathcal{P}}$ retains, with high probability, the key property that we need.

**Lemma 6.** *For any fixed multicommodity flow $f = (f_1, \ldots, f_k)$ solution, with high probability, all the flowpaths of $f$ are contained in $\widehat{\mathcal{P}}$.*

*Proof:* Let $p_1, \ldots p_q$ be the decomposition of the flow $f$ into flowpaths. By definition, all $p_i$ are contained in the set $\mathcal{P}$. Furthermore, by standard flow decomposition argument we know that $q \leq km \leq n^4$. Let us focus now on some particular path $p_i$. Let $1 \leq t \leq n$ be the number of arcs in this path, and let $j^*$ be the smallest $j$ for which $t \leq 2^j$. The probability that $p_i \in \mathcal{P}(S_{j^*}, 2^{j^*}) \subseteq \widehat{\mathcal{P}}$ is exactly the probability that at least one vertex from $p$ is in $S_{j^*}$. Simple computation shows that the probability that none among $t + 1$ vertices of $p_i$ is in $\mathcal{P}(S_{j^*}, 2^{j^*})$ is at most

$$(1 - \frac{10\ln n}{2^{j^*}})^{t+1} \leq e^{-\frac{10(t+1)\ln n}{2^{j^*}}} \leq e^{-5\ln n} = n^{-5}.$$

Therefore, by union bounding over all $q \leq n^4$ paths $p_i$, we get that indeed $\{p_1, \ldots, p_q\} \subseteq \widehat{\mathcal{P}}$ with high probability. ∎

### 3.1 $(\delta, M_{\max}, M_{\min}, \mathcal{Q})$-ADSP data structure

Once we defined set $\widehat{\mathcal{P}}$, our goal is to devise an efficient way of maintaining the $(1+\delta)$-approximate shortest paths with respect to it. We start by formally defining our task.

**Definition 7.** *For any $\delta \geq 0$, $M_{\max} \geq 2M_{\min} > 0$ and a set of paths $\mathcal{Q} \subseteq \mathcal{P}$, let the $\delta$-approximate decremental $(M_{\max}, M_{\min}, \mathcal{Q})$-shortest path problem $((\delta, M_{\max}, M_{\min}, \mathcal{Q})$-ADSP, for short) be a problem in which one maintains a directed graph $G$ with a length function $l$ on its arcs that supports four operations (sometimes we will refer to these operations as requests):*

- *Distance$(u, v, \beta)$, for $u, v \in V$, and $\beta \in [M_{\min}, M_{\max}/2]$: let $d^*$ be the length of the shortest (with respect to $l$) $u$-$v$ path in $\mathcal{Q}$; if $d^* \leq 2\beta$ then the query returns a value $d$ such that $d \leq d^* + \delta\beta$. If $d^* > 2\beta$, the query may return either $d$ as above, or $\infty$;*

- *Increase$(e, \omega)$, for $e \in E$ and $\omega \geq 0$: increases the length $l(e)$ of the arc $e$ by $\omega$;*

- *Path$(u, v, \beta)$, for $u, v \in V$, and $\beta \in [M_{\min}, M_{\max}/2]$: returns a $u$-$v$ path of length at most Distance$(u, v, \beta)$, as long as Distance$(u, v, \beta) \neq \infty$;*

- *SSrcDist$(u, \beta)$ for $u \in V$, and $\beta \in [M_{\min}, M_{\max}/2]$: returns Distance$(u, v, \beta)$ for all $v \in V$.*

Intuitively, $\beta$ is our guess on the interval $[\beta, 2\beta]$ in which the length of the shortest path we are interested in is. We say that $\beta$ is $(u, v)$-*accurate* for given $(\delta, M_{\max}, M_{\min}, \mathcal{Q})$-ADSP data structure $R$ and $u, v \in V$, if the length $d^*$ of the shortest $u$-$v$ path in $\mathcal{Q}$ is at least $\beta$ and the data structure $R$ returns a finite value in response to Distance$(u, v, \beta)$ query. Note that if $\beta$ is $(u, v)$-accurate then the $\delta\beta$-additive error guarantee on the distance estimation supplied by $R$ in response to Distance$(u, v, \beta)$ query implies a $(1 + \delta)$ *multiplicative* error guarantee. Also, as long as $d^*$ is at least $M_{\min}$ (this will be always the case in our applications), we can employ binary search to ask $O(\log \log M_{\max}/M_{\min})$ Distance$(u, v, \cdot)$ queries and either realize that $d^*$ is bigger than $M_{\max}$, or find $0 \leq i \leq \lceil \log M_{\max}/2M_{\min} \rceil$ such that $\beta_i = \min\{2^i M_{\min}, M_{\max}/2\}$ is $(u, v)$-accurate.[5] Finally, it is worth emphasizing that we do not require that the paths returned in response to $\mathsf{Path}(\cdot, \cdot, \cdot)$ queries are from $\mathcal{Q}$ – all we insist on is just that all the suitable paths from $\mathcal{Q}$ are considered when the path to be returned is chosen.

In section 6, we describe how the ideas and tools from dynamic graph algorithms lead to an implementation of the $(\delta, M_{\max}, M_{\min}, \widehat{\mathcal{P}})$-ADSP data structure that is tailored to maintain the shortest paths from set $\widehat{\mathcal{P}}$ and whose performance is described in the following theorem. The theorem is proved in section 6.2.

**Theorem 8.** *For any $1 > \delta > 0$, $M_{\max} \geq 2M_{\min} > 0$, $(\delta, M_{\max}, M_{\min}, \widehat{\mathcal{P}})$-ADSP data structure can be maintained in total expected time $\widetilde{O}(mn \frac{\log M_{\max}/M_{\min}}{\delta})$ plus additional $O(1)$ per $\mathsf{Increase}(\cdot, \cdot)$ request in the processed sequence. Each $\mathsf{Distance}(\cdot, \cdot, \cdot)$ and $\mathsf{Path}(\cdot, \cdot, \cdot)$ query can be answered in $\widetilde{O}(n)$ time, and each $\mathsf{SSrcDist}(\cdot, \cdot)$ query in $\widetilde{O}(m)$ time.*

## 3.2  Solving the decremental dynamic all-pair shortest paths problem using the $(\delta, M_{\max}, M_{\min}, \widehat{\mathcal{P}})$-ADSP data structure

Interestingly, we can use the $(\delta, M_{\max}, M_{\min}, \widehat{\mathcal{P}})$-ADSP data structure construction from Theorem 8 to obtain a $(1 + \varepsilon)$-approximate solution to the oblivious decremental dynamic all-pair shortest path problem in directed graphs with rational arc lengths, where obliviousness means that the sequence of requests that we process does not depend on the randomness used in the solution.

**Theorem 9.** *For any $1 > \varepsilon > 0$, and $L \geq 1$ there exists a $(1 + \varepsilon)$-approximate Monte Carlo solution to the oblivious decremental dynamic all-pair shortest paths problem on directed graphs where arc lengths are rational numbers between 1 and $L$, that has total maintenance cost of $\widetilde{O}(mn \frac{\log L}{\varepsilon})$ plus additional $O(1)$ per increase of the length of any arc, and answers shortest path queries for any vertex pair in $\widetilde{O}(n(\log \log_{(1+\varepsilon)} L)(\log \log L))$ time.*

Note that even when we allow lengths to be quite large (e.g. polynomial in $n$), the maintenance cost of our solution is still similar to the one that Roditty and Zwick achieved in [27] for undirected graphs with small integer lengths. Unfortunately, our distance query time is $\widetilde{O}(n)$ instead of the $O(1)$ time obtained in [27]. So, the gain that we get over a naïve solution for the problem is that we are able to answer $((1 + \varepsilon)$-approximately) shortest path queries for any vertex pair in $\widetilde{O}(n)$ time, as opposed to the $O(m + n \log n)$ time required by Dijkstra's algorithm. The proof of the theorem can be found in Appendix A.

## 4  Maximum multicommodity flow

We proceed to develop a faster algorithm for the maximum multicommodity flow problem – later we will generalize it to solve the weighted version of the problem. As we indicated in the previous section, the basic idea behind our improvement is modification of Fleischer's algorithm (presented in Figure 4) to make it exploit the dependencies between the oracle subproblems. More precisely, instead of employing Dijkstra's algorithm each time, we answer the shortest-path questions by querying a $(\delta, M_{\max}, M_{\min}, \widehat{\mathcal{P}})$-ADSP data structure (as described in Theorem 8) that we maintain for appropriate choice of $\delta$, $M_{\max}$, and $M_{\min}$, and where $\widehat{\mathcal{P}}$ is a sparsification of $\mathcal{P}$, as described in Definition 5. However, the straight-forward implementation of this idea encounters some difficulties.

---

[5]The binary search just chooses $i$ in the middle of the current range of values in which the desired value of $i$ may lie (initially this range is $[0, \lceil \log M_{\max}/2M_{\min} \rceil]$), and if $\mathsf{Distance}(u, v, \beta_i)$ query returns $\infty$ then the left half of the range (including the $i$ queried) is dropped, otherwise the other half is dropped.

First, we have to justify the fact that while answering shortest-path queries we take into account mainly paths in $\widehat{\mathcal{P}}$, as opposed to the whole set $\mathcal{P}$. Second, an important feature of Dijkstra's algorithm that is exploited in Fleischer's approximation scheme, is the fact that whenever one computes the distance between a pair of vertices using this algorithm it simultaneously computes all single source shortest-path distances. Unfortunately, in our case we cannot afford to replicate this approach; thus we need to circumvent this issue in a more careful manner. We address these problems below.

## 4.1 Existence of short paths in $\widehat{\mathcal{P}}$

As mentioned in section 2, the key ingredient used in the Garg-Könemann framework to bound the quality of the solution for the maximum multicommodity flow problem is the fact that for any length function $l$ of $G$ there is always a path $p$ in $\mathcal{P}$ with length $l(p)$ being at most $\sum_e l(e)u(e)/OPT$. We prove now that with high probability the same property still holds when we consider only paths in $\widehat{\mathcal{P}}$.

**Lemma 10.** *With high probability, for any length function $l$, there exists a path $p \in \widehat{\mathcal{P}}$ with $l(p) \leq \frac{\sum_e l(e)u(e)}{OPT}$, where $OPT$ is the optimal value of the maximum multicommodity flow.*

*Proof:* Let $f^* = (f_1^*, f_2^*, \ldots, f_k^*)$ be some optimal multicommodity flow with $\sum_i |f_i^*| = OPT$. By Lemma 6 we know that with high probability $\widehat{\mathcal{P}}$ contains all the flowpaths $p_1, \ldots, p_q$ of $f^*$. The fact that $f^*$ has to obey the capacity constraints implies that $\sum_e l(e)u(e) \geq \sum_{j=1}^q l(p_j)f^*(p_j)$. But $OPT = \sum_i |f_i^*| = \sum_{j=1}^q f^*(p_j)$; an averaging argument shows that there exists a $j^*$ such that $l(p_{j^*}) \leq \sum_e l(e)u(e)/OPT$ as desired. ∎

To get a slightly different perspective on the above statement, note that the only property of $\widehat{\mathcal{P}}$ we are using in the proof is that it contains (with high probability) all the flowpaths of some optimal solution. This means that if we consider a restriction $LP'$ of the linear program (1) in which we set to zero all the variables $f_p$ with $p \in \mathcal{P} \setminus \widehat{\mathcal{P}}$, then with high probability the optimum of this restricted linear program $LP'$ is still equal to the original optimum i.e. $OPT$. Therefore, one may view the statement of the above lemma as a simple consequence of the weak duality between $LP'$ and its dual linear program.

## 4.2 Randomized cycling through commodities

For a given value of $\widehat{\alpha}$, and some $(\delta, M_{\max}, M_{\min}, \widehat{\mathcal{P}})$-ADSP data structure $R$, we say that a source-sink pair $(s, t)$ is *admissible for $\widehat{\alpha}$ (with respect to $R$)* if upon querying $R$ with $\mathsf{Distance}(s, t, \widehat{\alpha})$ the obtained answer is at most $(1 + 2\delta)\widehat{\alpha}$. In other words, $(s, t)$ is admissible for $\widehat{\alpha}$ if $R$'s estimate of the distance from $s$ to $t$ in $\widehat{\mathcal{P}}$ is small enough that our algorithm could choose to augment the flow along this path – provided $\widehat{\alpha}$ was its current lower-bound estimate of the length of the shortest path in $\widehat{\mathcal{P}}$. Obviously, our algorithm is vitally interested in finding source-sink pairs that are admissible for its current value of $\widehat{\alpha}$ – these pairs are the ones that allow augmentation of the flow.

Unfortunately, given the set of all possible pairs $\{(s_1, t_1), \ldots, (s_k, t_k)\}$, and the data structure $R$, it is not clear at all which one among them (if any) are admissible for given $\widehat{\alpha}$. Note, however, that if we deem some source-sink pair $(s, t)$ inadmissible for $\widehat{\alpha}$ (by querying $R$ for the corresponding *s-t* distance) then, since our length function is always increasing, this pair will never become admissible for $\widehat{\alpha}$ again. This suggests the following natural approach for identification of admissible pairs for a given $\widehat{\alpha}$. We cycle through all the sink-source pairs and query $R$ for the corresponding distance, we stick with one pair as long as it is admissible, and move on once it becomes inadmissible. Clearly, the moment we cycled through all pairs, we know that all the pairs are inadmissible for $\widehat{\alpha}$ with respect to the current length function $l$. The problem with this approach is that the resulting number of $s$-$t$ distance queries is at least $k$ and thus this would lead to the somewhat prohibitive bottleneck of $\widetilde{\Omega}(kn)$ in the running time (note that $k$ can be $\Omega(n^2)$).

To alleviate this problem we note that a very similar issue arose in Fleischer's algorithm – she avoided the above bottleneck by grouping the source-sink pairs according to common sources and exploiting the fact that Dijkstra's algorithm computes all single source shortest-path distances simultaneously. Therefore, one execution of Dijkstra's algorithm either finds an (analog of) admissible pair, or deems *all* the pairs sharing the same source inadmissible. Unfortunately, although our $(\delta, M_{\max}, M_{\min}, \widehat{\mathcal{P}})$-ADSP data structure allows single source shortest-path distance queries, these queries require $\widetilde{O}(m)$ time and we cannot afford to use them to obtain $s$-$t$ distances in the manner Fleischer did – this could cause $\Omega(m^2 \varepsilon^{-2})$ worst-case running time. We therefore devise a different method of circumventing the bottleneck. To describe it, let us assume that we are given some vertex $s$, and a set $I(s)$ of source-sink pairs that have $s$ as their common source and that have not yet been deemed inadmissible for our current value of $\widehat{\alpha}$. Our procedure samples $\lceil \log n \rceil$ source-sink pairs from $I(s)$ and checks whether any of them is admissible using the $\mathsf{Distance}(\cdot, \cdot, \cdot)$ query. If so, then we return the admissible pair found. Otherwise, i.e. if none of them was admissible, we use the $\mathsf{SSrcDist}(\cdot, \cdot)$ query to check which (if any) source-sink pairs in $I(s)$ are inadmissible, remove them from the set $I(s)$ and return an admissible pair (if any was found). We repeat the whole procedure – if $I(s)$ became empty, we proceed to the next possible source $s$ – until we examine all source-sink pairs. The algorithm is summarized in Figure 6 – for convenience, we substituted for $\delta$, $M_{\max}$, and $M_{\min}$ the actual values used in our algorithm. The intuition behind this procedure is that if all $\lceil \log n \rceil$ samples from $I(s)$ turned out to be inadmissible then with probability at least $(1 - \frac{1}{n})$, at least half of the pairs in $I(s)$ is inadmissible, and therefore the $\mathsf{SSrcDist}(\cdot, \cdot)$ query will reduce the size of $I(s)$ by at least half. Therefore, as we will show later, the expected number of $\mathsf{SSrcDist}(\cdot, \cdot)$ queries is not too large.

## 4.3 Our algorithm

We present our algorithm for the maximum multicommodity flow problem in Figure 5. As we already noted, the basic idea behind it is making Fleischer's algorithm (presented in Figure 4) to answer distance queries using an $(\varepsilon/2, 1, \gamma, \widehat{\mathcal{P}})$-ADSP data structure (where $\widehat{\mathcal{P}}$ is constructed as in Definition 5). To implement this idea efficiently, we incorporated the randomized cycling through commodities described above.

We prove first that the flow produced by our algorithm is indeed close to the optimal one.

**Lemma 11.** *For any $0.15 > \varepsilon > 0$, with high probability, the algorithm presented in Figure 5 finds an $(1 - 3\varepsilon)$-approximate solution to the maximum multicommodity flow problem.*

*Proof:* Let $l_j$ be the length function $l$ after $j$-th augmentation of flow $f$, and let $\alpha(j)$ be the length of the shortest path in $\widehat{\mathcal{P}}$ with respect to $l_j$. Also, let $g_j$ be the total value of flow routed after $j$-th augmentation. Finally, let $OPT$ be the value of the optimal solution to our instance of the maximum multicommodity flow problem.

**Input** : Graph $G = (V, E)$, capacities $u(e)$, commodity pairs $\{(s_i, t_i)\}_i$, accuracy parameter $\varepsilon > 0$
**Output**: Flow $f$

Initialize $f \leftarrow \emptyset$, $\widehat{\alpha} \leftarrow \gamma$, and $l(e) \leftarrow \gamma$ for all arcs $e \in E$, where $\gamma = (1 + \varepsilon)/((1 + \varepsilon)n)^{1/\theta}$, and $\theta = \varepsilon(1 + \varepsilon)$
Sample sets $\{S_j\}_{j=1,\dots,\lceil \log n \rceil}$ to indicate the set $\widehat{\mathcal{P}}$ (see Definition 5)
Initialize $(\varepsilon/2, 1, \gamma, \widehat{\mathcal{P}})$-ADSP data structure $R$ as in Theorem 8
**while** $\widehat{\alpha} < 1$ **do**

    $I(s) \leftarrow \{i \mid s_i = s\}$ for each $s \in V$            (∗ initializing pairs to be examined ∗)
    **foreach** $s \in V$ with $I(s) \neq \emptyset$ **do**            (∗ cycling through commodities ∗)
        $\langle (s, t), I(s) \rangle \leftarrow$ Find_Admissible_Pair$(\varepsilon, R, \widehat{\alpha}, I(s))$
        **while** $I(v) \neq \emptyset$ **do**
            $p \leftarrow R.\mathsf{Path}(s, t, \widehat{\alpha})$
            Find the bottleneck capacity $u$ of $p$            (∗ $u \leftarrow \min_{e \in p} u(e)$ ∗)
            Augment the flow $f$ by routing $u$ units of flow along the path $p$
            **foreach** arc $e$ in $p$ **do** $R.\mathsf{Increase}(e, \frac{\varepsilon u l(e)}{u(e)})$; $l(e) \leftarrow l(e)(1 + \frac{\varepsilon u}{u(e)})$
            $\langle (s, t), I(s) \rangle \leftarrow$ Find_Admissible_Pair$(\varepsilon, R, \widehat{\alpha}, I(s))$
        **end**
    **end**
    $\widehat{\alpha} \leftarrow (1 + \varepsilon/2)\widehat{\alpha}$            (∗ increase the value of $\widehat{\alpha}$ ∗)
**end**
Scale $f$ down by a factor of $\max_{e \in E} f(e)/u(e)$ and return it

Figure 5: Improved algorithm for solving maximum multicommodity flow. Find_Admissible_Pair procedure is described in Figure 6.

---

**Input** : Accuracy parameter $\varepsilon > 0$, $(\varepsilon/2, 1, \gamma, \widehat{\mathcal{P}})$-ADSP data structure $R$, lower bound $\widehat{\alpha}$ on
        $\min_{p \in \widehat{\mathcal{P}}} l(p)$, and a set $I(s)$ of source-sink pairs that might be admissible for $\widehat{\alpha}$
**Output**: $\langle (s, t), I'(s) \rangle$, where $I'(s)$ is a subset of $I(s)$, and $(s, t)$ is: an admissible pair for $\widehat{\alpha}$ if
        $I'(s) \neq \emptyset$; an arbitrary pair otherwise

**for** $i = 1, \dots, \lceil \log n \rceil$ **do**
    Sample a random source-sink pair $(s, t)$ from $I(s)$
    **if** $R.Distance(s, t, \widehat{\alpha}) \leq (1 + \varepsilon)\widehat{\alpha}$ **then**            (∗ checking admissibility for $\widehat{\alpha}$ ∗)
        **return** $\langle (s, t), I(s) \rangle$
    **end**
**end**
                       (∗ No admissible pairs sampled ∗)
Use $R.\mathsf{SSrcDist}(s, \widehat{\alpha})$ to check admissibility for $\widehat{\alpha}$ of all the source-sink pairs in $I(s)$
Let $I'(s)$ be the subset of admissible pairs from $I(s)$
**if** $I'(s) \neq \emptyset$ **then**
    **return** $\langle (s, t), I'(s) \rangle$ where $(s, t) \in I'(s)$
**else**
    **return** $\langle (s, t), \emptyset \rangle$ where $(s, t)$ is an arbitrary pair
**end**

Figure 6: Procedure Find_Admissible_Pair

We start by proving that for every $j$, the value $\widehat{\alpha}_j$ of $\widehat{\alpha}$ immediately before $(j+1)$-th augmentation is at most $\alpha(j)$. To prove this, assume for the sake of contradiction that this is not the case; let $j^*$ be some $j$ for which $\widehat{\alpha}_j > \alpha(j)$. This means that there exists a $s$-$t$ path $p \in \widehat{\mathcal{P}}$ with $l_{j^*}(p) < \widehat{\alpha}_{j^*}$. Since $l_{j^*}(p) \geq \gamma$, it must have been the case that $\widehat{\alpha}_{j^*} > \gamma$ and the pair $(s,t)$ was deemed inadmissible for $\frac{\widehat{\alpha}_{j^*}}{(1+\varepsilon/2)}$ at some earlier point of the algorithm – otherwise the value of $\widehat{\alpha}$ would never increase up to $\widehat{\alpha}_{j^*}$. But this is impossible, since the length of $p$ could only increase over time and

$$l_{j^*}(p) + \frac{\varepsilon}{2}\frac{\widehat{\alpha}_{j^*}}{(1+\varepsilon/2)} < \widehat{\alpha}_{j^*} + \frac{\varepsilon}{2}\frac{\widehat{\alpha}_{j^*}}{(1+\varepsilon/2)} = (1+\varepsilon)\frac{\widehat{\alpha}_{j^*}}{(1+\varepsilon/2)}.$$

Thus $\mathsf{Distance}(s,t,\widehat{\alpha}_{j^*}/(1+\varepsilon/2))$ must have had return a value of at most $(1+\varepsilon)\frac{\widehat{\alpha}_{j^*}}{(1+\varepsilon/2)}$ which would deem the pair $(s,t)$ admissible.

Let us define $D(j) := \sum_{e \in E} l_j(e)u(e)$ to be the volume of $G$ with respect to $l_j$. For any $j \geq 1$ we have that $l_{j-1}(p_j) \leq (1+\varepsilon)\widehat{\alpha}_{j-1} \leq (1+\varepsilon)\alpha(j-1)$, and therefore

$$D(j) = \sum_e l_{j-1}(e)u(e) + \varepsilon \sum_{e \in p_j} l_{j-1}(e)(g_j - g_{j-1}) \leq D(j-1) + \theta\alpha(j-1)(g_j - g_{j-1}),$$

where $\theta = \varepsilon(1+\varepsilon)$. This in turn implies that

$$D(j) \leq D(0) + \theta\sum_{j'=1}^{j}(g_{j'} - g_{j'-1})\alpha(j'-1) \tag{3}$$

By applying Lemma 10 to the length function $l_j - l_0$, we get that with high probability, for every $j \geq 1$

$$OPT \leq \frac{D(j) - D(0)}{\alpha(j) - \gamma n}, \tag{4}$$

where we used the fact that $l_0(p) \leq \gamma n$ for any path $p$. Substituting (3) in equation (4) gives

$$\alpha(j) \leq \gamma n + \frac{\theta}{OPT}\sum_{j'=1}^{j}(g_{j'} - g_{j'-1})\alpha(j-1).$$

Now, if we focus on a particular $j$, the right hand side is maximized by setting all $\alpha(j')$ to their maximum possible values for all $0 \leq j' < j$. Let $\alpha'(j)$ be such a maximum value for any $j$. Thus we have

$$\alpha(j) \leq \alpha'(j) = \alpha'(j-1)(1 + \frac{\theta(g_j - g_{j-1})}{OPT}) \leq \alpha'(j-1)e^{\theta(g_j - g_{j-1})/OPT}$$

Since $\alpha'(0) = \gamma n$, this implies that $\alpha(j) \leq \gamma n e^{\theta g_j/OPT}$, and since we stop augmentations once $\alpha(j) \geq \widehat{\alpha}$ is at least 1, we know that after the final, say $j_f$-th, augmentation, we have

$$1 \leq \alpha(j_f) \leq \gamma n e^{\theta g_t/OPT}$$

and thus

$$\frac{OPT}{g_{j_f}} \leq \frac{\theta}{\ln(\gamma n)^{-1}}. \tag{5}$$

Since no arc can have length $l(e)$ larger than $(1+\varepsilon)$ throughout the algorithm, and routing $u(e)$ units of flow through $e$ causes increase in $l(e)$ by a factor of at least $(1+\varepsilon)$, the factor by which we have to scale down the flow $f$ at the end is at most $\lfloor \log_{1+\varepsilon} \frac{1+\varepsilon}{\gamma} \rfloor$. Therefore, by substituting the bound from (5), we see that the final value of flow $f$ is at least

$$|f| \geq \frac{g_{j_f}}{\lfloor \log_{1+\varepsilon} \frac{1+\varepsilon}{\gamma} \rfloor} \geq \frac{\ln(\gamma n)^{-1} OPT}{\theta \log_{1+\varepsilon} \frac{1+\varepsilon}{\gamma}}$$

Setting $\gamma = (1+\varepsilon)((1+\varepsilon)n)^{-1/\theta}$ and using the definition of $\theta$, we have

$$|f| \geq ((\frac{1}{\varepsilon(1+\varepsilon)} - 1)\ln(1+\varepsilon))OPT.$$

This quantity is at least $(1-3\varepsilon)OPT$ thereby concluding the proof. ∎

We are ready to establish the following theorem.

**Theorem 12.** *For any $0.15 > \varepsilon > 0$, with high probability, the algorithm presented in Figure 5 finds a $(1-3\varepsilon)$-approximate solution to the maximum multicommodity flow problem in expected $\widetilde{O}(mn\varepsilon^{-2})$ time.*

*Proof:* The fact that with high probability the final flow is a $(1-3\varepsilon)$-approximation to the optimal flow follows from Lemma 11.

To bound the running time of the algorithm, we note that it is dominated by the total cost of maintaining our $(\varepsilon/2, 1, \gamma, \widehat{\mathcal{P}})$-ADSP data structure $R$, and servicing all the requests issued to it – all other operations performed by the algorithm can be amortized within this total cost.

By Theorem 8, the total expected maintenance cost of our data structure $R$ is at most $\widetilde{O}(mn\frac{\log 1/\gamma}{\varepsilon}) = \widetilde{O}(mn\varepsilon^{-2})$. Also, note that each augmentation of the flow results in at least one (bottlenecking) arc having its length increased by a factor of $(1+\varepsilon)$, and during the algorithm no arc $e$ can have its length bigger than $(1+\varepsilon)$. Therefore, there can be at most $m\lfloor \log_{(1+\varepsilon)} \frac{(1+\varepsilon)}{\gamma} \rfloor$ flow augmentations. As a result, the cost of serving all $\mathsf{Path}(\cdot, \cdot, \cdot)$ and $\mathsf{Increase}(\cdot, \cdot)$ requests is at most $m\lfloor \log_{(1+\varepsilon)} \frac{(1+\varepsilon)}{\gamma} \rfloor(\widetilde{O}(n) + n \cdot O(1)) = \widetilde{O}(mn\varepsilon^{-2})$, as desired.

Now, to bound the time needed to service all the $\mathsf{Distance}(\cdot, \cdot, \cdot)$ queries, we note that there can be at most $m\lfloor \log_{(1+\varepsilon)} \frac{(1+\varepsilon)}{\gamma} \rfloor$ samplings of source-sink pairs in the *Find_Admissible_Pair* procedure that yield an admissible pair. This is so, since finding an admissible pair results in flow augmentation. Thus the total cost of samplings that found some admissible pair is at most $\lceil \log n \rceil m\lfloor \log_{(1+\varepsilon)} \frac{(1+\varepsilon)}{\gamma} \rfloor\widetilde{O}(n) = \widetilde{O}(mn\varepsilon^{-2})$. On the other hand, in cases when sampling does not yield an admissible path the time needed to serve all the corresponding $\mathsf{Distance}(\cdot, \cdot, \cdot)$ queries can be amortized in the time needed to serve $\mathsf{SSrcDist}(\cdot, \cdot)$ that is always issued afterward.

Therefore, all that is left to do is to bound the expected service cost of $\mathsf{SSrcDist}(\cdot, \cdot)$ queries. We call a $\mathsf{SSrcDist}(s, \widehat{\alpha})$ *successful* if it resulted in decreasing the size of $I(s)$ by at least a half. Note that there can be at most $\lceil \log n \rceil$ successful $\mathsf{SSrcDist}(\cdot, \cdot)$ queries per one source and one value of $\widehat{\alpha}$. As a result, the total time spent on answering successful queries is at most $\lfloor \log_{(1+\varepsilon/2)} \frac{1}{\gamma} \rfloor n \cdot \widetilde{O}(m) = \widetilde{O}(mn\varepsilon^{-2})$, since we have at most $\lfloor \log_{(1+\varepsilon/2)} \frac{1}{\gamma} \rfloor$ different values of $\widehat{\alpha}$. On the other hand, to bound the time taken by serving unsuccessful $\mathsf{SSrcDist}(\cdot, \cdot)$ queries, we notice that each (successful or not) $\mathsf{SSrcDist}(\cdot, \cdot)$ query either empties one set $I(s)$ for given source $s$ and a value of $\widehat{\alpha}$, or finds an admissible pair (which results in flow augmentation), therefore there can be at most $(n \lfloor \log_{(1+\varepsilon/2)} \frac{1}{\gamma} \rfloor + m \lfloor \log_{(1+\varepsilon)} \frac{(1+\varepsilon)}{\gamma} \rfloor)$ $\mathsf{SSrcDist}(\cdot, \cdot)$ queries in total. Moreover, the probability that a particular query is unsuccessful is at most $\frac{1}{n}$ - this follows from the fact that if $I(s)$ has more than a half of pairs admissible for given $\widehat{\alpha}$ then the probability that none among $\lceil \log n \rceil$ independent samples turns out to be admissible is at most $\left(\frac{1}{2}\right)^{\lceil \log n \rceil} \leq \frac{1}{n}$. Therefore, the total expected cost of this type of queries is at most $\frac{(n \lfloor \log_{(1+\varepsilon/2)} \frac{1}{\gamma} \rfloor + m \lfloor \log_{(1+\varepsilon)} \frac{(1+\varepsilon)}{\gamma} \rfloor)}{n} \cdot \widetilde{O}(m) = \widetilde{O}(\frac{m^2}{n\varepsilon^2}) = \widetilde{O}(mn\varepsilon^{-2})$, as desired. The theorem follows. ∎

## 4.4 Extension to weighted maximum multicommodity flow

Recall that the weighted maximum multicommodity flow problem is a generalization of the maximum multicommodity flow problem in which each commodity $i$ has a positive weight $w_i$ associated with it and we want to find a solution $f = (f_1, \ldots, f_k)$ that maximizes $\sum_i w_i |f_i|$. By scaling, we may ensure that the minimum weight is one, and maximum weight has some value $W$. In Appendix B we present a simple modification of our algorithm for the maximum multicommodity flow problem that can solve this generalization. We prove there the following statement.

**Corollary 13.** *For any $0.15 > \varepsilon > 0$, there exists a Monte Carlo algorithm that finds a $(1 - 3\varepsilon)$-approximate solution to weighted maximum multicommodity flow problem in expected $\widetilde{O}(mn\varepsilon^{-2} \log^2 W)$ time.*

# 5 Maximum concurrent flow

In the maximum concurrent flow problem, in addition to graph $G = (V, E)$ with capacities $u(\cdot)$ (we assume that $\min_e u(e) = 1$, and let $U := \max_e u(e)$), and $k$ source-sink pairs $(s_i, t_i)$, we also have a demand $d_i > 0$ associated with each commodity $i$. The task is to find a feasible multicommodity flow routing $\lambda d_i$ units of each commodity $i$ for maximum $\lambda$.

Let $\lambda^*$ be the optimal maximum concurrent flow value. When $\lambda^*$ is at least one, Garg and Könemann [12] describe an approximation scheme that runs in $\widetilde{O}((m + k)m\varepsilon^{-2})$ time – we will describe this algorithm shortly. To handle the cases when $\lambda^*$ is less than one, they describe a procedure to scale the demands so that it becomes at least one. This procedure computes, for each source-sink pair $(s_i, t_i)$ the value $\eta_i$ of maximum $s_i$-$t_i$ flow. Let $\eta^* = \min_i \eta_i/d_i$. Clearly, $\lambda^* \leq \eta^*$. Also, the fact that we can simultaneously route $1/k$ fraction of each of these $k$ maximum flows implies that $\lambda^* \geq \eta^*/k$. Therefore, scaling all the demands by $\eta^*/k$ ensures that $1 \leq \lambda^* \leq k$, and thus the approach for the case of $\lambda^* \geq 1$ can be applied. Subsequently, Fleischer [11] presented a different scaling procedure – based on maximum bottleneck path computations – that allows obtaining a $m$-approximation of $\lambda^*$ in time $\widetilde{O}(\min\{n, k\}m)$. This change made the preprocessing needed to ensure that $\lambda^* \geq 1$ no longer dominate the running time of the whole algorithm, thus the total running time of $\widetilde{O}((m + k)m\varepsilon^{-2})$ has been obtained. In [17] Karakostas further improved the running time of this algorithm by making it run in $\widetilde{O}((m^2 + kn)\varepsilon^{-2})$ time.

We proceed to a more detailed description of the algorithm of [12] for the case of $1 \leq \lambda^* \leq m$ – as we already said, the scaling procedure from [11] justifies this assumption. We start with a zero flow $- f = \emptyset$, and each arc $e$ has initially length $l(e) = \frac{\gamma}{u(e)}$, where $\gamma = (\frac{m}{1-\varepsilon})^{-1/\varepsilon}$. The algorithm proceeds in *phases* – each one of them consists of $k$ iterations. In iteration $i$, the algorithm tries to route $d_i$ units of flow from $s_i$ to $t_i$. This is done by repeating the following steps. First, a shortest (with respect to current length function $l$) $s_i$-$t_i$ path $p$ is found. Next, we compute the bottleneck capacity $u$, which is the minimum of the bottleneck capacity of the path $p$ and the remaining demand $\widehat{d_i}$ to be routed. We augment the flow by routing $u$ units of it along the path $p$. Finally, we increase the length of each arc $e$ of the path $p$ by a factor of $(1 + \frac{\varepsilon u}{u(e)})$, and decrease $\widehat{d_i}$ by $u$. The entire procedure stops when $D(l) := \sum_e l(e)u(e)$ – the volume of $G$ with respect to $l$, is at least one. The summary of the algorithm can be found in Figure 7.

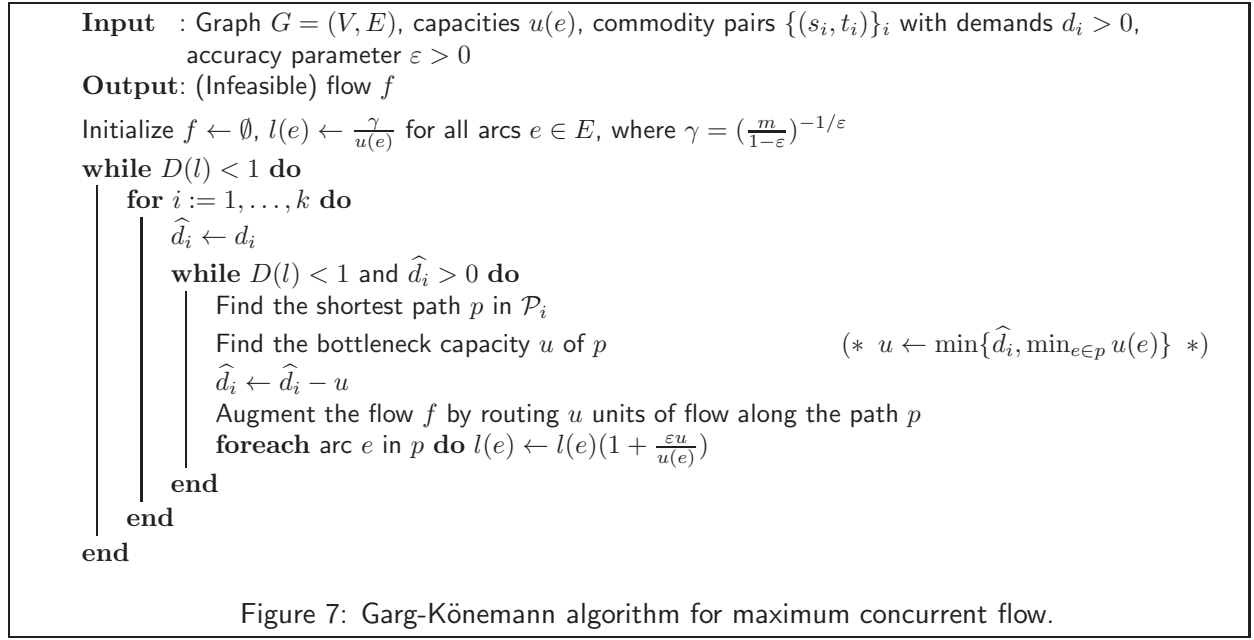

Figure 7: Garg-Könemann algorithm for maximum concurrent flow.

Garg and Könemann prove the following lemmas.

**Lemma 14.** *If $\lambda^* \geq 1$, the algorithm in Figure 7 terminates after at most $t := 1 + \lambda^* \log_{1+\varepsilon} 1/\gamma = 1 + \frac{\lambda^*}{\varepsilon} \log_{1+\varepsilon} \frac{m}{1-\varepsilon}$ phases.*

**Lemma 15.** *After $t-1$ phases, the algorithm has routed $(t-1)d_i$ units of each commodity $i$. Scaling the final flow by $\log_{1+\varepsilon} 1/\gamma$ yields a feasible flow of value $\lambda = \frac{t-1}{\log_{1+\varepsilon} 1/\gamma}$.*

**Lemma 16.** *If $\lambda^* \geq 1$, then the final flow scaled down by $\log_{1+\varepsilon} 1/\gamma$ is feasible and has a value of at least $(1 - 3\varepsilon)\lambda^*$.*

As we ensured that $1 \leq \lambda^* \leq m$, it is easy to see that the above lemmas imply that the algorithm in Figure 7, after at most $1 + \lambda^* \log_{1+\varepsilon} 1/\gamma \leq 1 + m \log_{1+\varepsilon} 1/\gamma$ phases, returns a $(1-3\varepsilon)$-approximation for the maximum concurrent flow. Unfortunately, the bound of $1 + m \log_{1+\varepsilon} 1/\gamma$ on the number of phases is not sufficient to obtain the $\widetilde{O}((m+k)m\varepsilon^{-2})$ running time (in fact, it only establishes the time bound of $\widetilde{O}((m+km)m\varepsilon^{-2})$). To reduce this dependence of the number of phases on $\lambda^*$, Garg and Könemann use a halving technique developed in [24]. They run the algorithm and if it does not stop after $T := 2 \log_{1+\varepsilon} 1/\gamma + 1$ phases then, by Lemma 14, it must be that $\lambda^* > 2$. In this case, they multiply the demands by two, so $\lambda^*$ is halved and still at least one. Next, they continue the algorithm and keep doubling the demands again if it does not stop after $T$ phases. Clearly, since initially $\lambda^* \leq m$, after repeating such doubling at most $\log m$ times the algorithm stops, and thus the total number of phases is at most $T \log m$. The number of phases can be reduce further to $O(T)$ by first applying the above approach with constant $\varepsilon$ to get a constant-factor approximation for $\lambda^*$ – this takes $O(\log^2 m)$ phases – and then with at most $O(T)$ more phases the $(1-3\varepsilon)$-approximation is obtained. Having established this bound on the number of phases, the bound of $\widetilde{O}((m+k)m\varepsilon^{-2})$ on the running time of the whole algorithm follows easily.

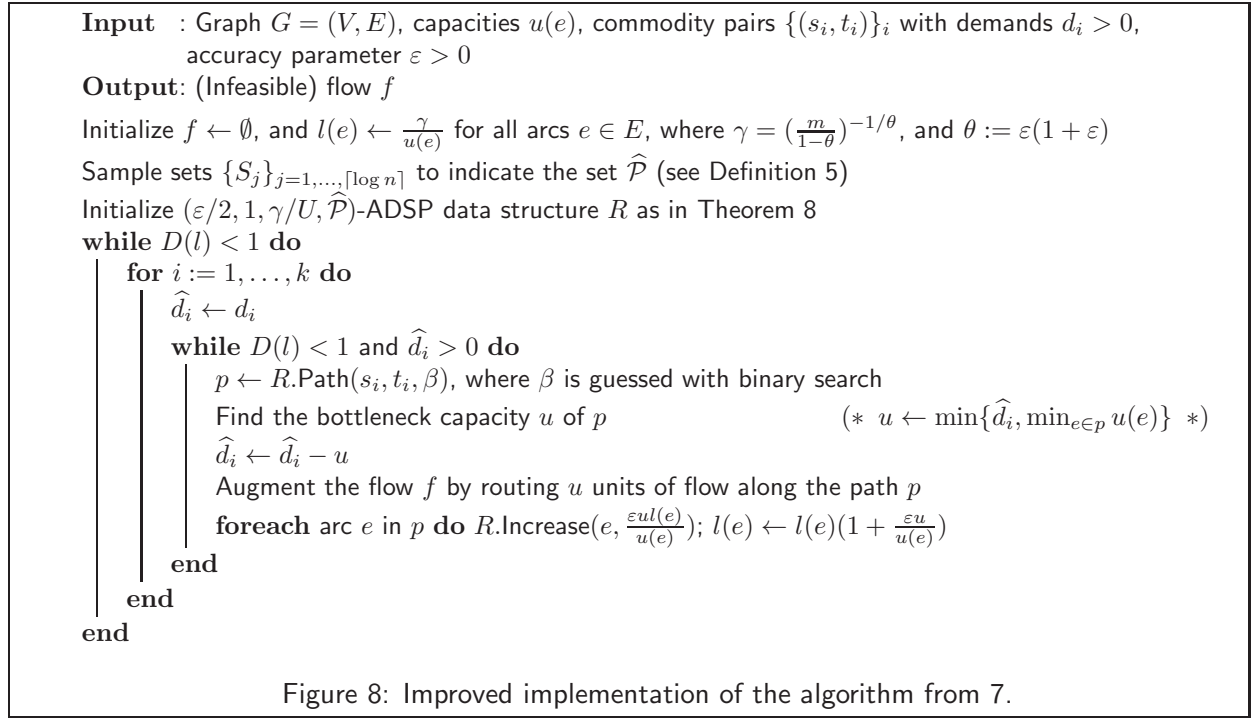

**Input** : Graph $G = (V, E)$, capacities $u(e)$, commodity pairs $\{(s_i, t_i)\}_i$ with demands $d_i > 0$,
  accuracy parameter $\varepsilon > 0$
**Output**: (Infeasible) flow $f$

Initialize $f \leftarrow \emptyset$, and $l(e) \leftarrow \frac{\gamma}{u(e)}$ for all arcs $e \in E$, where $\gamma = (\frac{m}{1-\theta})^{-1/\theta}$, and $\theta := \varepsilon(1+\varepsilon)$

Sample sets $\{S_j\}_{j=1,\dots,\lceil \log n \rceil}$ to indicate the set $\widehat{\mathcal{P}}$ (see Definition 5)

Initialize $(\varepsilon/2, 1, \gamma/U, \widehat{\mathcal{P}})$-ADSP data structure $R$ as in Theorem 8

**while** $D(l) < 1$ **do**
  **for** $i := 1, \dots, k$ **do**
    $\widehat{d_i} \leftarrow d_i$
    **while** $D(l) < 1$ and $\widehat{d_i} > 0$ **do**
      $p \leftarrow R.\mathsf{Path}(s_i, t_i, \beta)$, where $\beta$ is guessed with binary search
      Find the bottleneck capacity $u$ of $p$         ($* \ u \leftarrow \min\{\widehat{d_i}, \min_{e \in p} u(e)\} \ *$)
      $\widehat{d_i} \leftarrow \widehat{d_i} - u$
      Augment the flow $f$ by routing $u$ units of flow along the path $p$
      **foreach** arc $e$ in $p$ **do** $R.\mathsf{Increase}(e, \frac{\varepsilon u l(e)}{u(e)}); \ l(e) \leftarrow l(e)(1 + \frac{\varepsilon u}{u(e)})$
    **end**
  **end**
**end**

Figure 8: Improved implementation of the algorithm from 7.

## 5.1 Our algorithm

We present more efficient approximation scheme for maximum concurrent flow problem. Our improvement is based on making the algorithm from Figure 7 find the (approximately) shortest paths using $(\varepsilon, 1, \gamma/U, \widehat{\mathcal{P}})$-ADSP data structure instead of Dijkstra's algorithm, while using the same scaling procedures for $\lambda^*$ as above. Our new implementation of procedure from Figure 7 is presented in Figure 8. Note that whenever the algorithm issues $\mathsf{Path}(u, v, \cdot)$ request it uses binary search to find the $(u, v)$-accurate $\beta$ that yields small enough error – see discussion after Definition 7 for more details.

Let us define $\alpha(l) := \sum_i d_i \mathrm{dist}_i(l)$, where $\mathrm{dist}_i(l)$ is the length of the shortest $s_i$-$t_i$ path in $\widehat{\mathcal{P}}$ with respect to length function $l$. As it was the case for maximum multicommodity flow problem, we need to justify the fact that we focus our attention on shortest paths in $\widehat{\mathcal{P}}$ instead of the whole $\mathcal{P}$. As before, the following lemma could be also seen as a consequence of weak duality among appropriate primal and dual formulations of the problem.

**Lemma 17.** *With high probability, for any length function $l$, $\alpha(l) \leq D(l)/\lambda^*$, where $D(l) := \sum_e l(e)u(e)$.*

*Proof:* Similarly to the proof of Lemma 10, let us fix the optimal solution $f^* = (f_1^*, \ldots, f_k^*)$ that achieves the ratio of $\lambda^*$. By Lemma 6, with high probability, each $p_j$ belongs to $\widehat{\mathcal{P}} \cap \mathcal{P}_{i(j)}$, where $p_1, \ldots, p_q$ are the flowpaths of $f^*$ and $i(j)$ is the corresponding commodity. Now, since $f^*$ obeys capacity constraints we have that

$$D(l) = \sum_e l(e)u(e) \geq \sum_j l(p_j)f_{i(j)}^*(p_j) \geq \sum_{i=1}^k \lambda^* d_i \min_{p \in \mathcal{P}_i} l(p) \geq \lambda^* \alpha(l)$$

and the lemma follows.                                                                                       ∎

We are ready to prove the following theorem.

**Theorem 18.** *For any $0.15 > \varepsilon > 0$, there exists a Monte Carlo algorithm that finds a $(1-4\varepsilon)$-approximate solution for maximum concurrent flow problem in expected time $\widetilde{O}((m+k)n\varepsilon^{-2} \log U)$.*

*Proof:*

As it was already mentioned, we will be using the algorithm of Garg and Könemann with Fleischer's improvement (as described above), in which we use the procedure from Figure 8 instead of the one presented in Figure 7. To bound the running time of this new algorithm, we recall that the preprocessing that ensures that $1 \leq \lambda^* \leq m$ can be performed in $\widetilde{O}(\min\{k,n\}m)$ time. Also, we use the halving technique to ensure that the running time of the rest of the algorithm is bounded by the time needed to execute $O(T) = O(\log_{1+\varepsilon} 1/\gamma)$ phases of the procedure in Figure 8.

Now, it is easy to see that this running time is dominated by the cost of maintaining the data structure $R$, and the time needed to answer the queries $\mathsf{Path}(\cdot, \cdot, \cdot)$ (note there is at most $O(n)$ $\mathsf{Increase}(\cdot, \cdot)$ request per each $\mathsf{Path}(\cdot, \cdot, \cdot)$ query). By Theorem 8 we know that the maintenance cost is $\widetilde{O}(mn\frac{\log U/\gamma}{\varepsilon}) = \widetilde{O}(mn\varepsilon^{-1}(\varepsilon^{-1} + \log U)) = \widetilde{O}(mn\varepsilon^{-2}\log U)$. To upperbound the time needed to answer $\mathsf{Path}(\cdot, \cdot, \cdot)$ queries, we note that each such query results in augmentation of the flow, and each augmentation of the flow results either in increasing the length of at least one arc by $(1+\varepsilon)$, or it is the last augmentation for given commodity. But no arc can have length bigger than $(1+\varepsilon)/u(e)$, because $l(e)u(e) \leq D(l)$ and we stop when $D(l) \geq 1$, thus the total number of augmentations of the flow is at most $m\lfloor \log_{(1+\varepsilon)} \frac{(1+\varepsilon)}{\gamma} \rfloor + k \cdot O(T) = \widetilde{O}(m\varepsilon^{-2} + kT)$, which results in $\widetilde{O}((m+k)n\varepsilon^{-2}\log\log U/\gamma) = \widetilde{O}((m+k)n\varepsilon^{-2}\log\log U)$ needed to answer all $\mathsf{Path}(\cdot, \cdot, \cdot)$ queries (including the time needed to perform each binary search for $\beta$). As a result, the algorithm works in $\widetilde{O}((m+k)n\varepsilon^{-2}\log U)$ time, as desired.

We proceed now to lowerbounding the value of the final flow $f$ computed by our algorithm after scaling it down by the maximal congestion of the arcs. To this end, let, for given $j$, and $1 \leq q \leq q_j$, $f_{j,q}$ be the flow of commodity $i(j,q)$ that was routed along $s_{i(j,q)}$-$t_{i(j,q)}$ path $p_{j,q}$ in $q$-th augmentation of the flow $f$ in phase $j$, where $q_j$ is the total number of flow augmentations during phase $j$. Let $l_{j,q}$ be the length function $l$ after routing the flow $f_{j,q}$. For any $j$ and $q \geq 1$, the fact that we always find the $(s_{i(j,q)}, t_{i(j,q)})$-accurate $\beta$ for the $\mathsf{Path}(s_{i(j,q)}, t_{i(j,q)}, \beta)$ query, implies that $l_{j,q-1}(p_{j,q}) \leq (1+\varepsilon)\mathrm{dist}_{i(j,q)}(l_{j,q-1})$. Therefore, we have that

$$D(l_{j,q}) \leq D(l_{j,q-1}) + \theta \mathrm{dist}_{i(j,q)}(l_{j,q-1}),$$

where $\theta := \varepsilon(1+\varepsilon)$.

As a result, since the lengths of arcs can only increase, and in each completed phase we route exactly $d_i$ units of commodity $i$, for all $j \geq 1$ we have that

$$D(j) \leq D(j-1) + \theta \sum_i d_i \text{dist}_i(l_{j,q_j}) = D(j-1) + \theta \alpha(l_{j,q_j}),$$

where $D(j)$ denotes $D(l_{j,q_j})$.

By Lemma 17 we know that with high probability $\frac{D(j)}{\alpha(l_{j,q_j})} \geq \lambda^*$ for all $j$, thus we have that

$$D(j) \leq \frac{D(j-1)}{1 - \theta/\lambda^*}.$$

But, $D(0) = \sum_e \frac{\gamma}{u(e)} u(e) = m\gamma$, so for $j \geq 1$

$$
\begin{aligned}
D(j) &\leq \frac{m\gamma}{(1 - \theta/\lambda^*)^j} \\
&= \frac{m\gamma}{(1 - \theta/\lambda^*)}(1 + \frac{\theta}{(\lambda^* - \theta)})^{j-1} \\
&\leq \frac{m\gamma}{(1 - \theta/\lambda^*)}e^{\frac{\theta(j-1)}{\lambda^* - \theta}} \\
&\leq \frac{m\gamma}{(1 - \theta)}e^{\frac{\theta(j-1)}{\lambda^*(1-\theta)}},
\end{aligned}
$$

where the last inequality uses the fact that $\lambda^* \geq 1$.

The algorithm stops at the first phase $j_f$ during which $D(l) = \sum_e l(e)u(e) \geq 1$. Thus,

$$1 \leq D(j_f) \leq \frac{m\gamma}{(1-\theta)}e^{\frac{\theta(j_f-1)}{\lambda^*(1-\theta)}},$$

which in turn implies

$$j_f - 1 \geq \frac{\lambda^*(1-\theta)\ln\frac{1-\theta}{m\gamma}}{\theta}$$

Since we had $j_f - 1$ successfully completed phases, the flow produced by our procedure routes at least $(j_f - 1)d_i$ units of each commodity $i$. Unfortunately, this flow may be not feasible – it may violate some capacity constraints. But in our algorithm the length $l(e)$ of any arc $e$ cannot be bigger than $(1 + \varepsilon)/u(e)$. Thus, the fact that each arc starts with length $l(e) := \gamma/u(e)$ and each time a full unit of flow is routed through it, its length increases by a factor of at least $(1 + \varepsilon)$, implies that the congestion incurred at $e$ can be at most $\lfloor \log_{1+\varepsilon}(1+\varepsilon)/\gamma \rfloor$. Therefore, we see that the ratio $\lambda$ achieved by the final flow after scaling it down is at least

$$\lambda \geq \frac{j_f - 1}{\lfloor \log_{1+\varepsilon}(1+\varepsilon)/\gamma \rfloor} \geq \frac{\lambda^*(1-\theta)\ln\frac{1-\theta}{m\gamma}}{\theta \log_{1+\varepsilon} 1/\gamma}.$$

Plugging in $\gamma = (m/(1-\theta))^{-1/\theta}$ and unwinding the definition of $\theta$ yields

$$\lambda \geq \frac{(1 - \varepsilon(1+\varepsilon))^2 \ln(1+\varepsilon)}{\varepsilon(1+\varepsilon)}\lambda^* \geq (1 - 4\varepsilon)\lambda^*,$$

which proves that the flow is indeed a $(1 - 4\varepsilon)$-approximation of the maximum concurrent flow. ∎

## 5.2 Minimum cost concurrent flow

As noted in [12] and [11], the above approach for solving maximum concurrent flow extends easily to the case of the minimum concurrent flow problem in which we additionally have a cost function $c(\cdot)$ on arcs (routing one unit of flow along arc $e$ incurs a cost of $c(e)$), and we are interested in finding maximum concurrent flow whose total cost is within some target budget $B$. In fact, as pointed out in [11], this extension can be easily adapted to handle multiple budgets $B_j$ corresponding to different cost functions $c_j$.

To make our algorithm for the maximum concurrent flow problem handle the budget constraint, we introduce a dual variable $\phi$ corresponding to it. Initially, $\phi = \gamma/B$. Now, we modify our procedure in Figure 8 by stopping it once $D(l, \phi) := \sum_e l(e)u(e) + \phi B$ becomes at least one, and making it look for approximately shortest path with respect to the length function $l + \phi c$ instead of $l$ – we do this just by maintaining an $(\varepsilon, \gamma/(B+U), 1, \widehat{\mathcal{P}})$-ADSP data structure with respect to this new length function. Moreover, whenever some appropriate approximately shortest path $p \in \mathcal{P}_i$, for some $i$, is found by the algorithm, we augment the flow by routing along $p$ the amount of flow being maximal value $u$ such that: it does not overflow the bottleneck capacity $\min_{e \in p} u(e)$ of $p$, it does not exceed the remaining amount $\widehat{d_i}$ to be routed for commodity $i$ in this phase, and the cost of the flow – which is $c(p) = \sum_{e \in p} c(e)$ per each unit routed – does not exceed the budget $B$. After augmentation, we update $\widehat{d_i}$ and the value of $l(e)$ for the arcs $e$ of $p$ in usual manner. Moreover, we would like to increase the value of $\phi$ by a factor of $(1 + \frac{\varepsilon u c(p)}{B})$. Note however that each increase in $\phi$ results in increasing the length of all the arcs, thus we cannot afford to do the arc length updates each time $\phi$ changes. Instead, we do *aggregate* updates each time the value of $\phi$ grows by a factor of at least $(1 + \varepsilon)$ since the time the last aggregate updates were issued. Clearly, such update policy reduces the number of arc length increases to only $m \lceil \log_{(1+\varepsilon)} 1/\gamma \rceil = \widetilde{O}(m\varepsilon^{-2})$ while introducing an acceptable $(1 + \varepsilon)$ multiplicative error to our estimates of the length of the paths with respect to the length function $l + \phi c$. Finally, as it was the case for maximum concurrent flow problem, we have to ensure that the optimal ratio $\lambda^*$ for the budget-constrained version of the problem is at least one and not too large. We do this by a procedure presented in [11] that finds $\min\{k, n^2\}$-approximation to $\lambda^*$ in time $\widetilde{O}(\min\{n, k\}m)$.

Now, a reasoning completely analogous to the one for the maximum concurrent flow problem (see [12] for a highlight of straight-forward modifications in the proof) shows that, after proper scaling down at the end, we obtain a budget-constrained flow whose ratio is within $(1 - 5\varepsilon)$ of optimum[6], and the running time of the algorithm is $\widetilde{O}((m + k)n\varepsilon^{-2}\log(U + B))$. Therefore, we get the following corollary.

**Corollary 19.** *For any $0.15 > \varepsilon > 0$, there exists a Monte Carlo algorithm that finds a $(1 - 5\varepsilon)$-approximate solution for minimum cost concurrent flow problem in expected time $\widetilde{O}((m + k)n\varepsilon^{-2}\log(U + B))$.*

# 6 Construction of the $(\delta, M_{\max}, M_{\min}, \widehat{\mathcal{P}})$-ADSP data structure

In this section we describe a step-by-step construction of the $(\delta, M_{\max}, M_{\min}, \widehat{\mathcal{P}})$-ADSP data structure (defined in Definition 7) whose performance is described by Theorem 8. Let us fix our set $\widehat{\mathcal{P}} := \bigcup_{j=1}^{\lceil \log n \rceil} \mathcal{P}(S_j, 2^j)$, where $\{S_j\}_j$ are the sampled subsets of vertices as in Definition 5. Note that for each $j$ the expected size of $S_j$ is $O(n \log n / 2^j)$.

---

[6]We get $(1 - 5\varepsilon)$-approximation instead of $(1 - 4\varepsilon)$-approximation (as it was the case in Theorem 18), because of the additional $(1 + \varepsilon)$-multiplicative error introduced by our aggregate updates policy.

For given length function $l$, and real number $\rho$, let us define $l^{[\rho]}$ to be a length function with $l^{[\rho]}(e) = \lceil l(e)/\rho \rceil \rho$ i.e. $l^{[\rho]}$ corresponds to rounding-up the lengths of the arcs given by $l$ to the nearest multiple of $\rho$. The motivation behind this definition is captured in the following simple lemma.

**Lemma 20.** *For any $\rho > 0$, length function $l$, and $1 \leq j \leq \lceil \log n \rceil$, if there exists a path $p \in \mathcal{P}(S_j, 2^j)$ of length $l(p)$ then $l^{[\rho/2^j]}(p) \leq l(p) + \rho$.*

*Proof:* Consider a path $p \in \mathcal{P}(S_j, 2^j)$, for some $j$, by definition we have

$$l^{[\rho/2^j]}(p) = \sum_{e \in p} l^{[\rho/2^j]}(e) \leq \sum_{e \in p} (l(e) + \rho/2^j) \leq l(p) + \rho,$$

since $p$ can have at most $2^j$ arcs. ∎

As suggested by the above lemma, the basic idea behind our $(\delta, M_{\max}, M_{\min}, \widehat{\mathcal{P}})$-ADSP data structure construction is to maintain for each $j$ *exact* shortest paths from a set *larger* than $\mathcal{P}(S_j, 2^j)$ (namely, $\mathcal{P}(S_j)$), but with respect to the *rounded* version $l^{[\delta M_{\min}/2^j]}$ of the length function $l$, and to cap the length of these paths at $M_{\max} + \delta M_{\min}$. Note that we do not require our $(\delta, M_{\max}, M_{\min}, \widehat{\mathcal{P}})$-ADSP data structure to output paths from $\widehat{\mathcal{P}}$, thus this approach yields a correct solution. Moreover, as we show in section 6.1, using existing tools from dynamic graph algorithms we can obtain an implementation of this approach whose performance is close to the one postulated by Theorem 8, but with *linear* dependence of the maintenance cost on the ratio $\frac{M_{\max}}{M_{\min}}$ (as opposed to logarithmic one), and rather high service cost of $\mathsf{Increase}(\cdot, \cdot)$ queries. We alleviate these shortcomings in section 6.2, where we also prove Theorem 8.

## 6.1 Implementation of the $(\delta, M_{\max}, M_{\min}, \widehat{\mathcal{P}})$-ADSP with linear dependence on $\frac{M_{\max}}{M_{\min}}$

An important feature of the rounded length function $l^{[\rho]}$ for any $\rho > 0$, is that after dividing it by $\rho$ we obtain a length function that assigns integral lengths to arcs. Therefore, we are able to take advantage of existing tools for solving decremental shortest path problem in dynamic graphs with integer arc lengths. We start by defining this problem formally.

**Definition 21.** *For any integer $r \geq 0$ and a set of paths $\mathcal{Q} \subseteq \mathcal{P}$, let the decremental $(r, \mathcal{Q})$-shortest path problem ($(r, \mathcal{Q})$-DSPP for short) be a problem in which one maintains a directed graph $G$ with positive integral weights on its arcs, and that supports four operations:*

- $\overline{Distance}(u, v)$*, for $u, v \in V$: returns the length of the shortest $u$-$v$ path in $\mathcal{Q}$ if this length is at most $r$, and $\infty$ otherwise.*

- $\overline{Increase}(e, t)$*, for $e \in E$ and integer $t \geq 0$: increases the length of the arc $e$ by $t$*

- $\overline{Path}(u, v)$*, for $u, v \in V$: returns a $u$-$v$ path of length $\overline{Distance}(u, v)$, as long as $\overline{Distance}(u, v) \neq \infty$.*

- $\overline{SSrcDist}(u)$*, for $u \in V$: returns $\overline{Distance}(u, v)$ for all $v \in V$.*

We state first the following lemma which is just a simple and known extension of the classical construction of Even and Shiloach [10] (see also [29]).

**Lemma 22.** *For any $s \in V$ and positive integer $r$, $(r, \mathcal{P}(\{s\}))$-DSPP data structure can be maintained in total time $\widetilde{O}(mr)$ plus additional $O(\log n)$ per each $\overline{\mathsf{Increase}}(\cdot, \cdot)$ request. Each $\overline{\mathsf{Distance}}(\cdot, \cdot)$ query can be answered in $O(1)$ time and each $\overline{\mathsf{Path}}(\cdot, \cdot)$, and $\overline{\mathsf{SSrcDist}}(\cdot)$ query - in $O(n)$ time.*

For the sake of completeness, we prove the lemma in Appendix C.

We combine now the above construction of $(r, \mathcal{P}(\{s\}))$-DSPP data structure to obtain implementation of $(r, \mathcal{P}(U))$-DSPP.

**Lemma 23.** *For any $U \subseteq V$ and positive integer $r$, $(r, \mathcal{P}(U))$-DSPP data structure can be maintained in total time $\widetilde{O}(mr|U|)$ plus additional $\widetilde{O}(|U|)$ per each $\overline{\mathsf{Increase}}(\cdot, \cdot)$ request. Each $\overline{\mathsf{Distance}}(\cdot, \cdot)$ query can be answered in $O(|U|)$ time, each $\overline{\mathsf{Path}}(\cdot, \cdot)$ – in time $O(n)$, and each $\overline{\mathsf{SSrcDist}}(\cdot)$ query - in $O(m + n \log n)$ time.*

*Proof:* We maintain $(r, \mathcal{P}(\{s\}))$-DSPP data structure $R_s$ as in Lemma 22 for each $s \in U$. Clearly, the maintenance cost is $\widetilde{O}(mr|U|)$ plus additional $\widetilde{O}(|U|)$ per each $\overline{\mathsf{Increase}}(\cdot, \cdot)$ operation – we just forward each $\overline{\mathsf{Increase}}(\cdot, \cdot)$ operation to each $R_s$. Now, to serve $\overline{\mathsf{Distance}}(u, v)$ request we just issue $\overline{\mathsf{Distance}}(u, v)$ query to each $R_s$ that we maintain and return the answer yielding minimal value. Answering $\overline{\mathsf{Path}}(u, v)$ request consist of just querying each $R_s$ with $\overline{\mathsf{Distance}}(u, v)$, and forwarding $\overline{\mathsf{Path}}(u, v)$ request to $R_s$ returning the minimal distance. Finally, to serve $\overline{\mathsf{SSrcDist}}(u)$ query, we construct a graph $G_{u,U}$ that consists of $G$ equipped with current length function $l$, and additional vertex $u'$ from which there is arc to each $s \in U$ with length corresponding to the distance from $u$ to $s$ with respect to $l$. Note that construction of the graph $G_{u,U}$ can be performed in $O(m)$ time – in particular the length of each arc $(u', s)$ can be obtained by querying $R_s$ with $\overline{\mathsf{Distance}}(u, s)$. It is easy to see that if we compute single-source shortest path distances in $G_{u,U}$ from $u'$ to all $v \in V$ using Dijkstra's algorithm then we can obtain the value of $\overline{\mathsf{Distance}}(u, v)$ for each $v \in V$ by just returning the computed value if it is at most $r$ and $(u, v)$ is a source-sink pair (so, the corresponding $u$-$v$ path is in $\mathcal{P}$); and returning $\infty$ otherwise. Obviously, the total time required is $O(m + n \log n)$, as desired. ∎

We proceed now to designing an implementation of $(\delta, M_{\max}, M_{\min}, \widehat{\mathcal{P}})$-ADSP data structure that meets the time bounds of Theorem 8 except it maintenance time has linear – instead of logarithmic – dependence on $\frac{M_{\max}}{M_{\min}}$, and the time needed to serve $\mathsf{Increase}(\cdot, \cdot)$ request is much larger.

**Lemma 24.** *For any $\delta > 0$, and $M_{\max} > 2M_{\min} > 0$ we can maintain $(\delta, M_{\max}, M_{\min}, \widehat{\mathcal{P}})$-ADSP data structure in total expected time $\widetilde{O}(mn\frac{M_{\max}}{\delta M_{\min}})$ plus additional $\widetilde{O}(n)$ per each $\mathsf{Increase}(\cdot, \cdot)$ request. Each $\mathsf{Distance}(\cdot, \cdot, \cdot)$ and $\mathsf{Path}(\cdot, \cdot, \cdot)$ query can be answered in $\widetilde{O}(n)$ time, and each $\mathsf{SSrcDist}(\cdot, \cdot)$ query - in $\widetilde{O}(m)$ time.*

*Proof:* Let $l$ be the length function of our graph $G$. For each $1 \le j \le \lceil \log n \rceil$ we maintain a $(\lceil M_j/\rho_j \rceil, \mathcal{P}(S_j))$-DSPP data structure $R_j$ with respect to length function $l^j := l^{[\rho_j]}/\rho_j$, where $\rho_j := \delta M_{\min}/2^j$, and $M_j := M_{\max} + \delta M_{\min}$. Note that by definition $l^j$ is integral, so we are allowed to use the data structure from Lemma 23. Moreover, by Lemma 23 and Lemma 20 applied with $\rho = \rho_j 2^j$, we see that if there is an $s$-$t$ path $p$ in $\mathcal{P}(S_j, 2^j)$ of length $l(p) \in [M_{\min}, M_{\max}]$ then $R_j$ maintains a $s$-$t$ path $p' \in \mathcal{P}(S_j)$ whose length with respect to $l$ is at most $l(p) + 2^j \rho_j = l(p) + \delta M_{\min}$.

Now, to answer a $\mathsf{Distance}(u, v, \beta)$ query we just issue an $\overline{\mathsf{Distance}}(u, v)$ query to all $R_j$ and return the value that is minimal after multiplying it by the respective value of $\rho_j$. Note that we are ignoring the value of $\beta$ here - since $\beta \geq M_{\min}$, the accuracy of our answers is still sufficient. Similarly, as long as $\mathsf{Distance}(u, v, \beta) \neq \infty$, we answer $\mathsf{Path}(u, v, \beta)$ query by just returning the result of $\overline{\mathsf{Path}}(u, v)$ query forwarded to $R_j$ whose $\overline{\mathsf{Distance}}(u, v)$ (after multiplying by $\rho_j$) is minimal. We implement answering $\mathsf{SSrcDist}(u, \beta)$ query by just issuing $\overline{\mathsf{SSrcDist}}(u)$ queries to each $R_j$, and for each $v \in V$ we return the reported distance that is minimal (after multiplication by respective $\rho_j$). Finally, whenever there is a $\mathsf{Increase}(e, \omega)$ request, we increase the length function $l$ accordingly and issue $\overline{\mathsf{Increase}}(e, \lceil (l(e) + \omega)/\sigma_j \rceil - \lceil l(e)/\sigma_j \rceil)$ request to each $R_j$.

To analyze the performance of this implementation, we note that by Lemma 23 each $\mathsf{Distance}(\cdot, \cdot, \cdot)$ requires $O(\sum_j |S_j|) = \widetilde{O}(n)$ time, each $\mathsf{Path}(\cdot, \cdot, \cdot) - \widetilde{O}(n)$ time, and each $\mathsf{SSrcDist}(\cdot, \cdot) - \widetilde{O}(m)$ time. Also, the cost of serving $\mathsf{Increase}(e, \omega)$ request is $\widetilde{O}(\sum_j |S_j|) = \widetilde{O}(n)$. As a result, the total expected maintenance cost is, by Lemma 23:

$$\widetilde{O}\left( \sum_{j=1}^{\lceil \log n \rceil} m E[|S_j|] \lceil M_j/\rho_j \rceil \right) = \widetilde{O}\left( mn \sum_{j=1}^{\lceil \log n \rceil} \frac{2^j M_{\max}}{\delta M_{\min} 2^j} \right) = \widetilde{O}\left( mn \frac{M_{\max}}{\delta M_{\min}} \right),$$

where we used the fact that expected size of $S_j$ is $O(n \log n/2^j)$. The lemma follows. ∎

## 6.2 Proof of Theorem 8

As we mentioned in section 3, in our applications the ratio of $M_{\max}$ to $M_{\min}$ can be very large i.e. $\Omega(n^{1/\varepsilon})$ for $\varepsilon < 0.15$. Therefore, the linear dependence on this ratio of the maintenance time of the $(\delta, M_{\max}, M_{\min}, \widehat{\mathcal{P}})$-ADSP data structure from Lemma 24 is still prohibitive. To address this issue we refine our construction in the following lemma. Subsequently, we will deal with large service time of $\mathsf{Increase}(\cdot, \cdot)$ requests in the proof of Theorem 8.

**Lemma 25.** *For any $\delta > 0$, $M_{\max} \geq 2M_{\min} > 0$, $(\delta, M_{\max}, M_{\min}, \widehat{\mathcal{P}})$-ADSP data structure can be maintained in total expected time $\widetilde{O}(mn \frac{\log M_{\max}/M_{\min}}{\delta})$ plus additional $\widetilde{O}(n \log \frac{1}{\delta})$ per each $\mathsf{Increase}(\cdot, \cdot)$ request in the processed sequence. Each $\mathsf{Distance}(\cdot, \cdot, \cdot)$ and $\mathsf{Path}(\cdot, \cdot, \cdot)$ query can be answered in $\widetilde{O}(n)$ time, and each $\mathsf{SSrcDist}(\cdot, \cdot)$ query – in $\widetilde{O}(m)$ time.*

*Proof:*

For each $0 \leq b \leq \lfloor \log \frac{M_{\max}}{M_{\min}} \rfloor$, let us define $M_{\min}^b := 2^b M_{\min}$, and $M_{\max}^b = 2^{b+2} M_{\min}$. We will maintain for each $b$, a $(\delta, M_{\max}^b, M_{\min}^b, \widehat{\mathcal{P}})$-ADSP data structure $R_b$ as in Lemma 24. Intuitively, we divide the interval $[M_{\min}, M_{\max}]$ into exponentially growing and partially overlapping intervals $[M_{\min}^b, M_{\max}^b]$, and we will make each $R_b$ responsible for queries with $\beta$ falling into interval $[M_{\min}^b, M_{\max}^b/2]$.

More precisely, upon receiving $\mathsf{Distance}(u, v, \beta)$, $\mathsf{Path}(u, v, \beta)$, or $\mathsf{SSrcDist}(u, \beta)$ request, we just pass it to the unique $R_b$ with $M_{\min}^b \leq \beta \leq M_{\max}^b/2$, and report back the obtained answer. By Lemma 24 and definition of $(\delta, M_{\max}^b, M_{\min}^b, \widehat{\mathcal{P}})$-ADSP data structure, the supplied answer is correct, and the service cost is within desired bounds. Also, the part of the total expected maintenance cost that is independent of the number of $\mathsf{Increase}(\cdot, \cdot)$ requests is at most $\widetilde{O}(mn \frac{\log M_{\max}/M_{\min}}{\delta})$, as needed.

Therefore, it remains to design our way of handling $\mathsf{Increase}(\cdot, \cdot)$ requests, and bound the corresponding service cost. A straight-forward approach is to update the length function $l$ accordingly upon receiving $\mathsf{Increase}(e, \omega)$ request, and forward this request to all $R_b$. This would, however, result in $\widetilde{O}(n \log \frac{M_{\max}}{M_{\min}})$ service cost which is slightly suboptimal from our point of view. Our more refined implementation of handling $\mathsf{Increase}(e, \omega)$ request is based on two observations. First, we note that if at any point of time the length of arc $e$ increases to more than $2M_{\max}^b$ for some $b$, we can safely increase the length of this arc in $R_b$ to $\infty$ without violating the correctness of the answers supplied by $R_b$ – we call such event *deactivation of $e$ in $R_b$*. Second, we don't need to forward $\mathsf{Increase}(e, \omega)$ requests to $R_b$ for which $l(e) + \omega < \sigma_{\lceil \log n \rceil}^b$, where $\sigma_j^b = \frac{\delta M_{\min}^b}{2^j}$, and $l(e)$ is the current length of the arc $e$. This is so, since from Lemma 24 will have the rounded length $l^{[\sigma_j^b]}(e)$ of $e$ still equal to $\sigma_j^b$, for every $j$. Therefore, instead of passing to such $R_b$ $\mathsf{Increase}(e, \omega)$ requests each time they are issued, we just send an $\mathsf{Increase}(e, \omega')$ request to $R_b$ once the length of $e$ exceeds $\sigma_{\lceil \log n \rceil}^b$, where $\omega'$ is the total increase of the length of $e$ from the beginning up to the current value of $l(e)$ – we call such an event *activation of $e$ in $R_b$*.

In the light of the above, our handling of a $\mathsf{Increase}(e, \omega)$ request is as follows. Let $b_-$ be the largest $b$ with $2M_{\max}^b < l(e) + \omega$, and let $b_+$ be the largest $b$ with $l(e) + \omega \geq \sigma_{\lceil \log n \rceil}^b$. We start by deactivating $e$ in all $R_b$ with $b \leq b_-$ in which $e$ wasn't already deactivated, and activating $e$ (by increasing the length of $e$ to $l(e)$) in all $R_b$ with $b_- < b \leq b_+$ in which it wasn't activated yet. Next, we issue $\mathsf{Increase}(e, \omega)$ request to all $R_b$ with $b_- < b \leq b_+$, and we increase $l(e)$ by $\omega$. It is not hard to see that by the above two observations, this procedure does not violate the correctness of our implementation of $(\delta, M_{\max}, M_{\min}, \widehat{\mathcal{P}})$-ADSP data structure. Now, to bound the time needed to service $\mathsf{Increase}(\cdot, \cdot)$ request we note that each arc $e$ can be activated and deactivated in each $R_b$ at most once, and each such operation takes $\widetilde{O}(n)$ time. So, the total cost of these operation is at most $\widetilde{O}(mn \log \frac{M_{\max}}{M_{\min}})$ and this cost can be amortized within the total maintenance cost. To bound the time taken by processing the $\mathsf{Increase}(e, \omega)$ requests passed to all $R_b$ with $b_- < b \leq b_+$, we note that $l(e) + \omega < 2M_{\max}^{b-1} \leq 2^{j+3}\sigma_j^{b-1}/\delta$ for any $j$, thus $b_+ - b_- - 1 \leq \log \frac{2^{\lceil \log n \rceil + 3}}{\delta}$, and the total service time is at most $\widetilde{O}(n \log \frac{1}{\delta})$, as desired. The lemma follows. ∎

We are ready to prove Theorem 8.

*Proof:* [of Theorem 8] We maintain $(\delta/2, M_{\max}, M_{\min}, \widehat{\mathcal{P}})$-ADSP data structure $R$ as in Lemma 25. While serving the sequence of requests, we pass to $R$ all the $\mathsf{Path}(\cdot, \cdot, \cdot)$ requests and return back the answers supplied by $R$. Similarly, we pass $\mathsf{Distance}(\cdot, \cdot, \beta)$ and $\mathsf{SSrcDist}(\cdot, \beta)$ to $R$, and return back the values supplied by $R$ with $\frac{\delta\beta}{2}$ added to each of them. Finally, in case of $\mathsf{Increase}(\cdot, \cdot)$ requests we pass them to $R$ in an *aggregate* manner. Namely, in addition to $l$ – the (evolving) length function of the graph $G$ – we also maintain an *aggregated* length function $\hat{l}$. Initially, $\hat{l} = l$, and later, as $l$ evolves in an on-line manner, we increase $\hat{l}(e)$ to $l(e)$ for given arc $e$, each time $l(e)$ becomes greater than $\max\{(1 + \delta/8)\hat{l}(e), \delta M_{\min}/4n\}$. Note that this definition ensures that we have always $\hat{l}(e) \leq l(e) \leq (1 + \delta/8)\hat{l}(e) + \delta M_{\min}/4n$ for any arc $e$. Now, instead of passing to $R$ an $\mathsf{Increase}(e, \omega)$ request each time it is issued, we only issue an $\mathsf{Increase}(e, l(e) - \hat{l}(e))$ request to $R$ each time the value of $\hat{l}(e)$ increases to $l(e)$. In other words, we make $R$ to work with respect to the length function $\hat{l}$ instead of $l$.

Clearly, by Lemma 25 the time needed to answer $\mathsf{Distance}(\cdot, \cdot, \cdot)$, $\mathsf{Path}(\cdot, \cdot, \cdot)$, and $\mathsf{SSrcDist}(\cdot, \cdot)$ queries is within our intended bounds. Also, we can raise the length $l(e)$ of an arc $e$ to $\infty$ once its length exceeds $2M_{\max}$, without violating the correctness of our implementation of $(\delta, M_{\max}, M_{\min}, \widehat{\mathcal{P}})$-ADSP data structure. Thus, for given arc $e$, the length $\hat{l}(e)$ can increase at most $\lceil \log_{(1+\delta/8)} \frac{4M_{\max}n}{\delta M_{\min}} \rceil + 1 = \widetilde{O}(\frac{\log M_{\max}/M_{\min}}{\delta})$ times. Therefore, we issue at most $\widetilde{O}(m \frac{\log M_{\max}/M_{\min}}{\delta})$ $\mathsf{Increase}(\cdot, \cdot)$ requests to $R$, which by Lemma 25 takes at most $\widetilde{O}(mn \frac{\log M_{\max}/M_{\min}}{\delta})$ time to process. As a result, the total maintenance time of our construction is $\widetilde{O}(mn\frac{\log M_{\max}/M_{\min}}{\delta})$ plus additional $O(1)$ time per each Increase$(\cdot, \cdot)$ request in the sequence, as desired.

To prove that our construction is a correct implementation of $(\delta, M_{\max}, M_{\min}, \widehat{\mathcal{P}})$-ADSP data structure, consider some $s$-$t$ path $p$ in $\widehat{\mathcal{P}}$ whose length $l(p)$ is at most $2\beta$ for some $\beta \in [M_{\min}, M_{\max}/2]$. Now, upon being queried with Distance$(s, t, \beta)$ request, our data structure will return a value $d' = d + \delta\beta/2$, where $d$ is the value returned by $R$ as an answer to Distance$(s, t, \beta)$ request passed. Since $\hat{l}(p) \le l(p)$, we have $d' = d + \delta\beta/2 \le \hat{l}(p) + \delta\beta/2 + \delta\beta/2 \le l(p) + \delta\beta$, as desired. Moreover, upon Path$(s, t, \beta)$ query we return a path $p'$ with $\hat{l}(p') \le d$. This means that

$$l(p') \le \sum_{e \in p'}(1 + \delta/8)\hat{l}(e) + \delta M_{\min}/4n \le (1 + \delta/8)\hat{l}(p') + \delta\beta/4 \le (1 + \delta/8)d + \delta\beta/4 \le d + \delta\beta/2 = d',$$

since $d \le \hat{l}(p) + \delta\beta/2 \le 2\beta + \delta\beta/2 < 3\beta$ for $\delta < 1$. The theorem follows. ∎

# Acknowledgments

We are grateful to Michel Goemans, Debmalya Panigrahi, and anonymous reviewers for helpful comments. We also thank Lisa Fleischer, and David Shmoys for providing useful references.

# A    Proof of Theorem 9

*Proof:* [of Theorem 9] Let $G = (V, E)$ be the graph of our interest, and let $l_i$ be the length function of $G$ after processing $i$ requests from our sequence. Consider a $u$-$v$ path $p$, for some $u, v \in V$, that is the shortest $u$-$v$ path in $G$ with respect to $l_i$ for some $i$. Our construction is based on the following simple observation: for any $\delta > 0$, and for all $i' \geq i$, as long as for all arcs $e$ of $p$, $l_{i'}(e)$ is at most $(1 + \delta)l_i(e)$, $p$ remains to be a $(1 + \delta)$-approximate shortest $u$-$v$ path in $G$ with respect to $l_{i'}$. Note that we have $n^2$ different $(u, v)$ pairs, and each of $m$ arcs can increase its length by a factor $(1 + \delta)$ at most $\lceil \log_{(1+\delta)} L \rceil$ times. Therefore this observation implies that for any $\delta > 0$ there exists a set $\mathcal{Q}(\delta)$ of $O(mn^2 \frac{\log L}{\delta})$ paths such that for any $(u, v)$ and $i$ there exists a $u$-$v$ path $p$ in $\mathcal{Q}(\delta)$ that has length $l_i(p)$ within $(1 + \delta)$ of the length $l_i(p^*)$ of the shortest (with respect to $l_i$) $u$-$v$ path $p^*$ in $G$.

In the light of the above, our solution for the decremental all-pair shortest path problem is based on maintaining $(\varepsilon/3, Ln, 1, \widehat{\mathcal{Q}}(\varepsilon/3))$-ADSP data structure $R$, where $\widehat{\mathcal{Q}}(\varepsilon/3)$ is the set constructed as in Definition 5 after we make $\mathcal{P}$ to be the set of all paths in $G$ and change the sampling probabilities $p_j$ to $\min\{\frac{10\ln(n\lceil\log_{(1+\varepsilon/3)}L\rceil)}{2^j},1\}$ for $j=1,\ldots,\lceil\log n\rceil$. Note that by reasoning completely analogous to the one from the proof of Lemma 6, we can argue that with high probability $\mathcal{Q}(\varepsilon/3)\subseteq\widehat{\mathcal{Q}}(\varepsilon/3)$. Also, by straight-forward adjustment of the construction from Theorem 8 we obtain an implementation of $(\varepsilon/3,Ln,1,\widehat{\mathcal{Q}}(\varepsilon/3))$-ADSP data structure that has total maintenance cost of $\widetilde{O}(mn\frac{\log L}{\varepsilon})$ plus $O(1)$ time per each $\mathsf{Increase}(\cdot,\cdot)$ request, and that answers $\mathsf{Path}(\cdot,\cdot,\cdot)$ and $\mathsf{Distance}(\cdot,\cdot,\cdot)$ queries in $\widetilde{O}(n\log\log_{(1+\varepsilon)}L)$ time. Therefore, if we process the request sequence by just passing arc length increase requests to $R$, and answering each $u$-$v$ shortest path query by issuing to $R$ a $\mathsf{Path}(u,v,\beta)$ query – where $(u,v)$-accurate $\beta$ is found through binary search using $O(\log\log L)$ $\mathsf{Distance}(u,v,\cdot)$ queries (see discussion after Definition 7), then the Definition 7 ensures that we obtain a correct $(1+\varepsilon)$-approximate solution for decremental all-pair shortest path problem whose performance obeys the desired bounds. The theorem follows.  ∎

# B    Extension to weighted maximum multicommodity flow

One may obtain a linear programming formulation of weighted maximum multicommodity problem by simply changing the objective value of LP (1) to $\sum_{i=1}^{k}w_i\sum_{p\in\mathcal{P}_i}f_p$. Note that the dual of this modified LP is the LP (2) where constraint corresponding to $p\in\mathcal{P}_i$, for some $1\leq i\leq k$, becomes $\sum_{e\in p}l(e)\geq w_i$. This suggests that in the modified algorithm our criterion for selecting paths along which we augment the flow should be (approximate) minimization of the length of the path divided by its corresponding weight. As a result, we need to prove an analog of Lemma 10 that will certify that (with high probability) there is always a path in $\widehat{\mathcal{P}}$ whose ratio of length to the corresponding weight is small enough.

**Lemma 26.** *With high probability, for any length function $l$, and weights $w_1,\ldots,w_k$, there exists a $s_i$-$t_i$ path $p\in\widehat{\mathcal{P}}$, for some $1\leq i\leq k$, with $\frac{l(p)}{w_i}\leq\frac{\sum_e l(e)u(e)}{OPT}$, where $OPT$ is optimal value of the maximum weighted multicommodity flow.*

*Proof:* Let $f^*=(f_1^*,\ldots,f_k^*)$ be some optimal multicommodity flow with $\sum_i w_i|f_i^*|=OPT$. By Lemma 6 we know that with high probability $\widehat{\mathcal{P}}$ contains all the flowpaths $p_1,\ldots,p_q$ of $f^*$. The fact that $f^*$ has to obey the capacity constraints implies that $\sum_e l(e)u(e)\geq\sum_{j=1}^{q}l(p_j)f^*(p_j)$. But $OPT=\sum_i w_i|f_i^*|=\sum_{j=1}^{q}w_{i(j)}f^*(p_j)$, where $i(j)$ is such that $p_j\in\mathcal{P}_{i(j)}$. Therefore, simple averaging argument shows that there exists $j^*$ such that $\frac{l(p_{j^*})}{w_{i(j^*)}}\leq\frac{\sum_e l(e)u(e)}{OPT}$, as desired.  ∎

Similarly to the case of Lemma 10, the above lemma can also be seen as a consequence of weak duality between primal LP formulation of the problem, restricted to paths in $\widehat{\mathcal{P}}$, and corresponding dual linear program. We are ready to prove the corollary.

*Proof:* [of Corollary 13] We modify our algorithm in Figure 5 as follows. We decrease the value of $\gamma$ to $(1+\varepsilon)W/((1+\varepsilon)nW)^{1/\theta}$. Also, $\widehat{\alpha}$ will be now our lowerbound on $\min_{1\leq i\leq k}\min_{p\in\mathcal{P}_i}l(p)/w_i$. We still terminate the algorithm when $\widehat{\alpha}$ becomes at least one. This means that some of the paths we are interested in can have length up to $W$. As a result, we make $R$ to be a $(\varepsilon/2,W,\gamma,\widehat{\mathcal{P}})$-ADSP data structure (as opposed to $(\varepsilon/2,1,\gamma,\widehat{\mathcal{P}})$-ADSP). Also, for given source-sink pair $(s_i,t_i)$, the admissibility for $\widehat{\alpha}$ condition that is checked in procedure $\mathsf{Find\ Admissible\ Pair}$ becomes: $R$ upon querying with $\mathsf{Distance}(s_i,t_i,W_{j_i}\widehat{\alpha})$ query returns a value that is at most $(1+\varepsilon)w_i\widehat{\alpha}$, where $W_j=(3/2)^j$ for $j=1,\ldots,\lfloor\log_{3/2}W\rfloor$, and $W_{j_i}$ is the largest $W_j$ that is smaller than $w_i$. In particular, we change all the $\mathsf{Distance}(s_i,t_i,\widehat{\alpha})$ ($\mathsf{Path}(s_i,t_i,\widehat{\alpha})$ resp.) queries to $\mathsf{Distance}(s_i,t_i,W_{j_i}\widehat{\alpha})$

(Path$(s_i, t_i, W_{j_i} \widehat{\alpha})$ resp.). Also, whenever our previous algorithm was making SSrcDist$(s, \widehat{\alpha})$ call, we make it instead issue SSrcDist$(s, W_j \widehat{\alpha})$ calls for all $j = 1, \ldots, \lfloor \log_{3/2} W \rfloor$ and check admissibility of all the source-sink pairs in $I(s)$ using answers to these calls.

The running time analysis of our original algorithm from the proof of Theorem 12 can be easily adapted to show that the expected running time of the modified algorithm is $\widetilde{O}(mn\varepsilon^{-2} \log^2 W)$. The only differences are: the smaller value of $\gamma$, the fact that in Find Admissible Pair procedure we make $\lfloor \log_{3/2} W \rfloor$ SSrcDist$(\cdot, \cdot)$ queries instead of one, and that length of an arc can become as large as $(1+\varepsilon)W$ so there can be as many as $m \lfloor \log_{(1+\varepsilon)} \frac{(1+\varepsilon)W}{\gamma} \rfloor = \widetilde{O}(m\varepsilon^{-2} \log W)$ flow augmentations.

To prove the bound on the quality of returned solution, for $j \geq 1$, let $h_j - h_{j-1}$ be the increase in *weighted* flow value after $j$th augmentation of the flow along path $p_j \in \mathcal{P}_{i(j)}$, where $i(j)$ is the commodity to which $p_j$ corresponds, i.e. $h_j - h_{j-1} = w_{i(j)} u_j$, where $u_j$ is the amount of the flow we routed in $j$th augmentation, and $h_0 = 0$. Let $l_j$ be the length function $l$ after $j$th augmentation of the flow, and let $\alpha(j) = \min_{1 \leq i \leq k} \min_{p \in \mathcal{P}_i \cap \widehat{\mathcal{P}}} l_j(p) / w_i$. Finally, let $OPT$ be the value of optimal solution to our instance of the weighted maximum multicommodity flow problem.

By argument analogous to the one in the proof of Lemma 11, we can prove that $l_{j-1}(p_j) \leq (1+\varepsilon) w_{i(j)} \alpha(j-1)$. Now, if we define, as before, $D(j) := \sum_e l_j(e) u(e)$ to be the volume of $G$ with respect to $l_j$, then for $j \geq 1$ we have

$$
\begin{aligned}
D(j) &= \sum_e l_{j-1}(e) u(e) + \varepsilon \sum_{e \in p_j} l_{j-1}(e) u_j \\
&\leq D(j-1) + \theta w_{i(j)} \alpha(j-1)(h_j - h_{j-1}) / w_{i(j)} \\
&= D(j-1) + \theta \alpha(j-1)(h_j - h_{j-1}),
\end{aligned}
$$

where, as before, $\theta := \varepsilon(1+\varepsilon)$.

Now, we can apply Lemma 26 to length function $l_j - l_0$, to conclude that $OPT \leq \frac{D(j) - D(0)}{\alpha(j) - \gamma n}$ (cf. (4) in the proof of the Lemma 11). By continuing exactly as in Lemma 11 (and keeping in mind that maximum length of an arc can be $(1+\varepsilon)W$ now) we get the following bound on the value of the final feasible flow $h_f$:

$$
h_f \geq \frac{\ln(n\gamma)^{-1} OPT}{\theta \log_{(1+\varepsilon)} \frac{(1+\varepsilon)W}{\gamma}}.
$$

With our choice of $\gamma$, this is at least $(1 - 3\varepsilon)OPT$, as before. ∎

## C   Proof of Lemma 22

*Proof:* [of Lemma 22]

First, we notice that to prove the lemma it is sufficient to show that we can maintain, within desired time bounds, the single-source shortest paths tree of all $v$-$s$ paths that have length at most $r$, for any $v \in V$. Indeed, once we achieve this, it will also imply that we can keep the single-source shortest paths tree of all $s$-$v$ paths having length at most $r$, for any $v \in V$. Now, to implement $(r, \mathcal{P}(\{s\}))$-DSPP data structure we just maintain both single-source shortest path trees and whenever queried $\overline{\text{Distance}}(u, v)$ we answer $\overline{\text{Distance}}(u, s) + \overline{\text{Distance}}(s, v)$ if this sum does not exceed $r$ and $(u, v)$ is a source sink pair (i.e. the corresponding $u$-$v$ path is in $\mathcal{P}$); and $\infty$ otherwise.

Note that our trees allow finding $\overline{\mathsf{Distance}}(u, s)$ and $\overline{\mathsf{Distance}}(s, u)$ in $O(1)$ time. Similarly, we can answer queries $\overline{\mathsf{Path}}(u, v)$, as long as $\overline{\mathsf{Distance}}(u, s) + \overline{\mathsf{Distance}}(s, v) \leq r$, by just concatenating paths $\overline{\mathsf{Path}}(u, s)$ and $\overline{\mathsf{Path}}(s, v)$ that we can obtain from our trees in $O(n)$ time. Finally to answer $\overline{\mathsf{SSrcDist}}(u)$ query we just issue a $\overline{\mathsf{Distance}}(u, v)$ query – that we already know how to handle – for each $v \in V$. It is easy to see that all the running times will fit within the desired bounds as long as our maintenance of single-source shortest paths tree will obey these bounds.

Our way of maintaining such a single-source shortest path tree of all $v$-$s$ paths is a simple extension of the classical construction of Even and Shiloach [10] (see also [29]) who showed how to dynamically maintain in an *unweighted* directed graph, a decremental single-source shortest paths tree up to depth $d$, in $O(md)$ total running time.

In our data structure, each vertex $v$ will keep a variable $d[v]$ whose value will be always equal to $v$'s current distance to $s$. Moreover, each vertex $v$ keeps a priority queue $Q[v]$ of all its outgoing arcs, where the key of a particular arc $(v, u)$ is equal to the current value of $d[u] + l((v, u))$. We want the queue to support three operations: $Insert(e, t)$ that adds an arc to a queue with key equal to $t$, $FindMin$ - returns the arc with smallest value of the key, and $SetKey(e, t)$ that sets the value of the key of arc $e$ to $t$. By using e.g. Fibonacci heap implementation of such queue, $FindMin$ can be performed in $O(1)$ time, and each of the remaining operations can be done in $O(\log n)$ amortized time.[7]

The initialization of the data structure can be easily done by computing the single-source shortest path tree from $s$ using Dijkstra algorithm and inserting arcs into appropriate queues, which takes $O(m \log n)$ time. Also, $\overline{\mathsf{Distance}}(\cdot, s)$ queries can be easily answered in $O(1)$ time by just returning $d[v]$. Finally, the implementations of the $\overline{\mathsf{Path}}(\cdot, s)$, and $\overline{\mathsf{Increase}}(\cdot, \cdot)$ can be found in Figure 9. Clearly, answering $\overline{\mathsf{Path}}(\cdot, s)$ query takes at most $O(n)$ time. Now, the total time needed to serve $w$ $\overline{\mathsf{Increase}}(\cdot, \cdot)$ request is at most $O(\log n)$ times the total number of $Scan(\cdot)$ calls. But, since for a particular arc $e = (u, v)$ $Scan(e)$ is called only if either $\overline{\mathsf{Increase}}(e, \cdot)$ was called; or $d[v] < \infty$ and $d[v]$ increases by at least one, we see that this number is at most $m(r+1) + w$, which gives the desired running time.

∎

---

[7]Note that $SetKey(e, t)$ operation can be translated to 'standard' priority queue operations as: decrease the key of $e$ by a value of $\infty$, extract the minimal element, and insert $e$ again with new value of the key.

**Procedure** $\overline{\mathsf{Path}}(v, s)$:

  **if** $v = s$ **then**

    |   **return** an empty path $\emptyset$

  **else**

    |   $e = (v, u) \leftarrow Q[v].FindMin$

    |   **return** $\overline{\mathsf{Path}}(u, s) \cup \{e\}$

  **end**

**Procedure** $\overline{\mathsf{Increase}}(e, t)$:

  $l(e) \leftarrow l(e) + t$

  $Scan(e)$

**Procedure** $Scan(e)$ :

  $Q[u].SetKey(e, d[v] + l(e))$

  $f = (u, v') \leftarrow Q[u].FindMin$

  **if** $d[v'] + l(f) > r$ **then**

    |   $d[u] \leftarrow \infty$

  **end**

  **if** $d[v'] + l(f) > d[u]$ **then**

    |   $d[u] \leftarrow d[v'] + l(f)$

    |   **foreach** arc $f'$ incoming to $u$ **do** $Scan(f')$

  **end**

Figure 9: Implementation of procedures $\overline{\mathsf{Path}}(v, s)$ and $\overline{\mathsf{Increase}}(e, t)$, where $e = (u, v)$